\RequirePackage{lineno}
\documentclass[12pt,amsmath,amssymb]{article}

\usepackage{rsc}
\usepackage{graphicx}
\usepackage{dcolumn}
\usepackage{bm}
\usepackage{epsfig}
\usepackage{epstopdf}
\usepackage{amsmath}
\usepackage{amsfonts}
\usepackage{stmaryrd}
\usepackage{color}
\usepackage{lineno}
\usepackage{url}

\newcommand{\ud}{\,\mathrm{d}}

\begin{document}

\pagewiselinenumbers

\title{Impedance Spectra of Mixed Conductors: \\a 2D Study of Ceria}

\author{Francesco Ciucci,$^a$ \\ Yong Hao$^b$ \\ David G. Goodwin$^{a}$
 \\[3mm]
$^a$ California Institute of Technology \\
Department of Mechanical Engineering \\
Pasadena, CA 91125 USA. E-mail: francesco@alumni.caltech.edu\\[1mm]
$^b$ California Institute of Technology \\ 
Department of Materials Science, \\
Pasadena, CA 91125 USA. E-mail: haoyong@caltech.edu}

\date{\today}

\begin{abstract}
In this paper we  develop an analytical framework  for the study of electrochemical impedance of mixed ionic and electronic conductors (MIEC). The framework is based on first-principles and it features the coupling of electrochemical reactions, surface
transport and bulk transport processes. We utilize this work to analyze two-dimensional systems relevant for fuel cell science via finite element method (FEM). Alternate Current Impedance Spectroscopy (AC-IS or IS) of a ceria symmetric cell is simulated near equilibrium condition (zero bias) for a wide array of working conditions including variations of  temperature and $H_2$ partial pressure on a two-dimensional doped Ceria sample with patterned metal electrodes. The model shows agreement of IS curves with the experimental literature with the relative error on the impedance being consistently below 2\%. Important two-dimensional effects such as the effects of thickness decrease and the influence of variable electronic and ionic diffusivities on the impedance spectra are also explored.\end{abstract}

\maketitle

\section{\label{sec:Introduction}Introduction}
Mixed ionic and electronic conductors (MIEC), or  in short mixed conductors, are substances capable of conducting both electrons and ions, and for that reason they are used in many applications, most notably in catalysis and eletrochemistry: they have been employed in gas sensors, fuel cells, oxygen permeation membranes, oxygen pumps and electrolyzers. 

The study of the alternate current properties of MIEC aides in understanding many of the physical chemical phenomena related to the behavior of defects, electrochemistry and interfaces. A technique frequently used to probe the interplay between these processes is impedance spectroscopy (IS). IS consists in injecting a "small" sinusoidal current into an electrochemical sample, a fuel cell for example, which is initially under steady-state conditions. This perturbation in turn induces a small sinusoidal and de-phased perturbation of the voltage. From the measurements of voltage and current over a wide set of frequencies, one can compute the complex impedance of the system. When the experiment is compared against a suitable model, impedance spectroscopy helps understand the linear physics of electro-active materials. 

The tools used to deconvolute impedance spectra and relate them to physical-chemical quantities are usually limited to one-dimensional equivalent circuits \cite{ISI:000168035300017} \cite{ISI:000181439500003}. Even though the 1D approach is very useful because it enables the comparison of different processes, it sometimes fails to help interpret satisfactorily physical chemical phenomena that extend to several dimensions.
Only a handful of works attempted to scale up to two dimensions, and generally have been constrained to the steady-state setting \cite{mebane:A421} \cite{Fleig:July2004:1385-3449:637} \cite{Adler200035}.

In this paper we develop a fast method for the computation of impedance spectra for highly-doped mixed conductors in a 2D setting under geometrically symmetric conditions. The system studied was chosen so that it is not too cumbersome algebraically and readily relatable to experiments. However the methodology is general and it can be easily extended to 3D, to dissymmetric systems under non-zero bias and to complex chemical boundary conditions. 

The paper proceeds as follows: we first develop a model for impedance spectroscopy and determine the impedance equations \cite{macdonald:4982}, then we compare our results to experimental data, finally we study the influence of parameter variation on the IS: the thickness of the sample, the rates of the chemical reactions at the exposed MIEC surface and the diffusivity profiles.

After non-dimensionalization of the full drift diffusion equations, we find that the ratio between the Debye length and the characteristic length scale of the material is remarkably large, hence we singularly perturb the governing equations and we deduce that electroneutrality is satisfied for a large portion of the sample. Then we apply a small sinusoidal perturbation to the potential, which mathematically translates into a regular perturbation of the equations; after formal algebraic manipulations we collect first order terms and deduce two complex and linear partial differential equations in 2D space and time. Thank to linearity, the Fourier transformation of these equations and their boundary conditions leads to the determination of the complex impedance spectroscopy equations which we solve in 2D space for the frequencies of interest. 

We verify our numerical results against experiments that are relevant for fuel cell applications. In particular, we study the case of a Samarium Doped Ceria (SDC) sample, immersed in a uniform atmosphere of argon, hydrogen and water vapor. The sample is symmetric and reversible and has been the subject of extensive research \cite{ISI:000259269100064} \cite{ISI:000252830600012} \cite{ISI:000232773100001}.  We find excellent agreement between the computed impedance spectra and experimental data. This shows that the approximations and the model are likely to be valid, hence this framework could help address a number of important fundamental physical/chemical issues in mixed conductors.

\section{System Under Study}

\begin{figure}
\scalebox{0.5}{\includegraphics{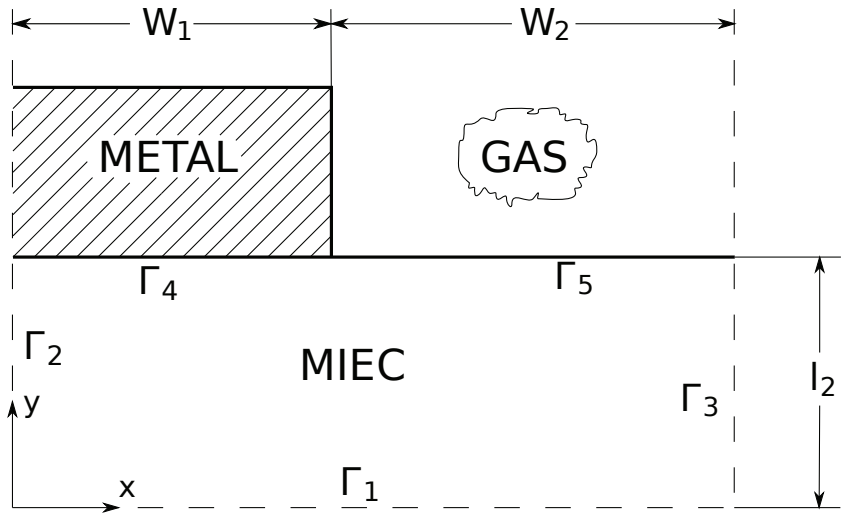}}
\caption{\label{fig:mat_domain} Schematic depiction of the domain under study with annotation of the boundary names and dimensions. The domain is composed by an MIEC slab of half-thickness $l_2$ which is mirror symmetric with respect to $\Gamma_1$. On top of the  slab there is a metal stripe infinitely long deposited over the surface $\Gamma_4$, the surface $\Gamma_5$ is exposed to the gas phase. The overall sample is mirror symmetric with respect to $\Gamma_2$ and $\Gamma_3$.}
\end{figure}

The physical system under study is a two-dimensional assembly which consists of a mixed oxygen ion and electron conductor slab of thickness $2l_2$ sandwiched between two identical patterned metal current collectors, Fig~\ref{fig:mat_domain}. The patterned collectors are repeated and symmetrical with respect to the  center line $\Gamma_1$. Hence the system to be reduced to a repeating cell using the mirror symmetry lines $\Gamma_1$, $\Gamma_2$ and $\Gamma_3$. All sides of the sample are placed in a uniform gas environment. Two charge-carrying species are considered: oxygen vacancies, denoted by the subscript ion, and electrons, denoted by eon. 

The framework we propose is very broad in scope, however we specialize our study to Samarium Doped Ceria (SDC). Doped ceria is a class of materials that has recently gained prominent relevance in fuel cell technology \cite{Gorte-00-Nature} \cite{Trovarelli-02}. We suppose that the uniform gas environment consists of a mixture of hydrogen and water vapor and we solve the electrochemical potential and current of both charge carriers using a linear and time-independent model, which we develop via perturbation techniques and Fourier transformation. We mainly compare our computational work to the data of Lai et al. \cite{ISI:000252830600012} but we also leverage on some results of Chueh et al. \cite{ISI:000259269100064} to justify the boundary conditions. Both works study SDC-15 (15\% samarium doping), hence the background dopant particles per unit volume, $B$,  is well defined and reported in Tab.~\ref{tab:geometric_values}.

The surface dimensions are kept constant: the width of the metal $|$ ceria interface ($\Gamma_4$) is $2W_1 = 3\mu m$ and the width of the gas $|$ ceria interface ($\Gamma_5$) is $2W_2 = 5 \mu m$.  The thickness of the MIEC is set to be $2l_2 = 1mm$, unless otherwise specified.  Due to high electronic mobility in the metal, the thickness of the metal stripe does not affect to the calculation, and thus the thickness of the electrolyte is in effect the thickness of the cell. Hence we assume that the characteristic length scale of the sample under study is $l_c = 10\mu m$. The data mentioned above is summarized in Tab.~\ref{tab:geometric_values}.

The assumptions of the model are rather standard for MIEC. We set that the gas $|$ metal $|$ ceria interface, or triple-phase boundary, has a negligible contribution compared to surface reactions \cite{Adler-96}. We further treat the surface chemistry as one global reaction, and do not consider diffusion of adsorbed species on the surface \cite{Kee-05}. Combined with the final assumption that the metal $|$ ceria interface is reversible to electrons, i.e., a Ohmic condition \cite{mebane:A421}, we are only considering two steps in the electrode reaction pathway, for instance, surface reactions at the active site of the SDC$|$Gas interface and electron drift-diffusion from the active site to the metal current collector both along the SDC|gas interface and through the SDC bulk.

We indicate the equilibrium quantities, such as electron and oxygen vacancy concentration, with the superscript $(0)$. In order to determine equilibrium concentrations of charge carriers, we consider the following gas phase and bulk defect reactions:
\begin{subequations}
\begin{eqnarray}
H_2({\rm gas})+O_2({\rm gas}) &\rightleftharpoons& H_2O({\rm gas}) \\
O_O^x  &\rightleftharpoons&V_O^{\bullet\bullet}+\frac{1}{2}O_2({\rm gas})+2e' 
\label{global_{eon}quilibrium_reactions}
\end{eqnarray}
\end{subequations}
where the Kr\"oger-Vink notation is used \cite{Kroger-56}, i.e. $V_O^{\bullet\bullet}$ is a vacant site in the crystal, $e'$ is an electron, and $O_O^x$ an oxygen site in the crystal (superscripts $\bullet$, $'$ and $x$ indicate respectively +1 charge,  -1 charge and zero charge). At equilibrium the number of vacant sites per unit volume is $c_{ion}^{(0)}$, and the number of electrons per unit volume is $c_{eon}^{(0)}$. At equilibrium the following two quantities will be constants:

\begin{subequations}
\begin{eqnarray}
K_g &=& \displaystyle\frac{\tilde p_{H_2O}^2}{\tilde p_{H_2}^2\tilde p_{O_2}} \label{eqn:Kg}\\
K_r &=& \displaystyle\left(\frac{c_{eon}^{(0)}}{B}\right)^2 \frac{c_{ion}^{(0)}}{B} \tilde p_{O_2}^{1/2}
\end{eqnarray}
\label{eqns_global_reactions}
\end{subequations}
\noindent in addition to that, electroneutrality will be satisfied throught the sample:
\begin{equation}
1+ \displaystyle \frac{c_{eon}^{(0)}}{B} - 2\frac{c_{ion}^{(0)}}{B}=0
\end{equation}

\noindent where $\tilde p_k =\displaystyle \frac{p_k}{1 {\rm atm}}$ and $p_k$ is the partial pressure of species $k$. In the dilute limit, at a given temperature and partial pressure, we solve for the equilibrium concentrations of vacancies $c_{ion}^{(0)}\approx B/2$ and electrons $c_{eon}^{(0)} \approx B \frac{\sqrt{2 Kr}}{[Sm_{Ce}']^{ 1.5} {\tilde p_{O_2}}^{0.25}}$.

Finally we assume that the mobilities $u$ of all species are given in Tab.~ \ref{tab:physic_params}, from Lai et al. \cite{ISI:000252830600012}.

\section{\label{sec:Math_Equation}Background}
\subsection{Asymptotic Modeling of Mixed Conduction in the Bulk}
A mixed conductor is a substance capable of conducting two or more charged species of opposite sign. Mass and charge transport in solids are described, at a mesoscopic level, by drift diffusion (DD) equations. The derivation of these equations is given in many textbooks, see for example \cite{Newman-04}. For clarity we will shortly rewrite them here. For a mobile species $m$, the continuity portion of the DD equations is expressed by equations of the form:

\begin{equation}
{\partial{c_m}\over{\partial t}}+\nabla\cdot{\bf{j}}_m^P={\dot{\omega}}_m
\label{carrier_conservation}
\end{equation}
\noindent where $c_m$ is the concentration of species $m$, ${\bf{j}}_m^P$ is the particle (superscript $P$) flux of species $m$ per unit area and ${\dot{\omega}}_m$ is its net rate of creation per unit volume.

We will assume the following phenomenological relationship for the flux of species $m$ (this relation is valid for $\nabla T \simeq 0$ and $\nabla P \simeq 0$ \cite{Prigogine-61}):

\begin{equation}
{\bf j}_m^P = -{c_m D_m\over k_b T} \nabla\tilde\mu_m
\end{equation}
\noindent where $D_m$ and $\tilde \mu_m$ are respectively its diffusivity, given by Einstein's relation $D_m = u_m k_b T/z_m$ ($u_m$ is the mobility), and its electrochemical potential, given by an expression of the type:
\begin{equation}
\tilde \mu_m = \mu^0_m +k_bT \log(c_m f_m(c_m, T,P)) + z_m e \phi
\label{eqn:electrochem_general}
\end{equation}

In the latter $e$ is the elementary charge and $\phi$ is the electric potential, $f_m$ is the activity of species $m$ and $z_m$ is its integer charge, i.e. -1 for electrons, +2 for oxygen vacancies in an oxide and  $\mu^0_m$ is a reference value. We also define the $\star$-electrochemical potential of a species $m$ as:
\begin{equation}
\tilde \mu_m^\star = \frac{\tilde \mu_m}{z_m}
\label{eqn:electrochem_star}
\end{equation}



The same equations are sometime expressed in a different way; if we define the conductivity ${\sigma_m} = {e^2 c_m D_m z_m^2 \over k_b T}$, we will deduce from Eqn.s~\ref{carrier_conservation} and~\ref{eqn:electrochem_general} that:



\begin{equation}
{ \partial{c_m}\over{\partial t} } -  \nabla \cdot \left \{ \left( D_m + {\partial \log f_m \over \partial \log c_m} \right) {\nabla c_m}  + {\sigma_m\over z_m e}\nabla \phi \right\} = \dot{\omega}_m
\label{carrier_conservation_mu}
\end{equation}


Here we suppose the presence of two mobile species: oxygen vacancies, which we indicate with the subscript $ion$ ($z_{ion}=+2$), and electrons, subscript $eon$ ($z_{eon}=-1$). The distribution of electrons and vacancies is thus described by 3 equations: one for the electric field (Poisson's equation for the potential) and two for the mobile species conservation. This set of equations can be written as:
\begin{subequations}
\begin{eqnarray}
\triangle \phi = \frac{e}{\varepsilon}\left( B+c_{eon}-2c_{ion}\right)\\
\partial_t c_{eon}+\nabla\cdot\left( -D_{eon}c_{eon}\nabla \frac{\tilde \mu_{eon}}{k_B T} \right) =0\\
\partial_t c_{ion}+\nabla\cdot\left( -D_{ion}c_{ion}\nabla \frac{\tilde \mu_{ion}}{k_B T} \right) =0
\end{eqnarray}
\label{eqn:fundamental}
\end{subequations}
where $\varepsilon$ is the permittivity of the medium, $B$ is the background dopant concentration in number of particles per unit volume and where we have chosen $\dot \omega_{eon} = \dot \omega_{ion}=0 $. 
In the dilute limit \cite{Adler-96} \cite{Adler-97} \cite{Liu-97} \cite{Liu-99} \cite{Riess-03}, one has:
\begin{subequations}
\begin{eqnarray}
\tilde \mu_{eon} = k_BT \log\left( \frac{c_{eon}}{c_{eon}^{0}}\right)-e\phi + \tilde \mu^{0}_{eon}\\
\tilde \mu_{ion} = k_BT \log \left( \frac{c_{ion}}{c_{ion}^{0}}\right)+2e\phi + \tilde \mu^{0}_{ion}
\end{eqnarray}
\end{subequations}
where $c_{ion}^0$ and $c_{eon}^0$ are reference values.

Non-dimensionalization of the Eqn.s~\ref{eqn:fundamental} with respect to its relevant parameters proves to be crucial in order to understand appropriate time and length scales. We apply the transformations: $(\mathbf x, t)\rightarrow(\mathbf{\tilde x}, \tilde t)$ such that $\mathbf x = l_c \mathbf{\tilde x}$ and $t= \tau\tilde t$. At this point we suppose the diffusivities $D_{eon}$ and $D_{ion}$ are uniform (we shall relax this approximation later). Also, we define $U_T ={k_b T}/{e}$, $\tilde \phi ={\phi}/{U_T}$,  $\tau_n = \displaystyle{l_c^2}/{D_{eon}}$,  $\tau_p = \displaystyle{l_c^2}/{D_{ion}}$ and  $\tau= \min(\tau_n, \tau_p)$. Obviously $\nabla_{x}(\cdot) = \frac{1}{l_c}\nabla_{\tilde x}(\cdot)$ and $\partial_{t}(\cdot) = \frac{1}{\tau}\partial_{\tilde t}(\cdot)$. So Eqn.~\ref{eqn:fundamental} becomes:
\begin{subequations}
\begin{eqnarray}
\displaystyle\triangle_{\tilde x} \tilde \phi = \frac{e l_c^2 B}{\varepsilon U_T}\left( 1+\frac{c_{eon}^{(0)}}{B}\frac{c_{eon}}{c_{eon}^{(0)}}-2\frac{c_{ion}^{(0)}}{B}\frac{c_{ion}}{c_{ion}^{(0)}}\right)
\label{eqn:fundamental_dilution_non_dim_poisson} \\
\displaystyle\frac{\tau_n}{\tau}\partial_{\tilde t} \frac{c_{eon}}{c_{eon}^{(0)}}+\nabla_{\tilde x}\cdot\left( -\nabla_{\tilde x} \frac{c_{eon}}{c_{eon}^{(0)}}+\frac{c_{eon}}{c_{eon}^{(0)}} \nabla_{\tilde x} \tilde \phi\right) =0 \label{eqn:fundamental_dilution_non_dim_eon} \\
\displaystyle\frac{\tau_p}{\tau}\partial_{\tilde t} \frac{c_{ion}}{c_{ion}^{(0)}}-\nabla_{\tilde x}\cdot\left( \nabla_{\tilde x} \frac{c_{ion}}{c_{ion}^{(0)}}+2 \frac{c_{ion}}{c_{ion}^{(0)}}\nabla_{\tilde x} \tilde \phi\right) =0 \label{eqn:fundamental_dilution_non_dim_ion}
\end{eqnarray}
\label{eqn:fundamental_dilution_non_dim}
\end{subequations}
\noindent where $c_{eon}^{(0)}$ and $c_{ion}^{(0)}$ are equilibrium values \cite{ISI:000232773100001}.
Define now the Debye length $\lambda_D =\sqrt{ \frac{\varepsilon U_T}{e  B}}$ and $\lambda = \frac{l_c}{\lambda_D}$. We suppose $\lambda\gg 1$, which holds true for highly doped MIECs and sufficiently large characteristic dimensions, and we use singular perturbation of Eqn.~\ref{eqn:fundamental_dilution_non_dim_poisson} to obtain \cite{Marchowich-90}:
\begin{equation}
1+\frac{c_{eon}^{(0)}}{B}\frac{c_{eon}}{c_{eon}^{(0)}}-2\frac{c_{ion}^{(0)}}{B}\frac{c_{ion}}{c_{ion}^{(0)}} =0
\label{eqn:electroneutrality_condition}
\end{equation}
In view of the latter, we can drop Eqn.~\ref{eqn:fundamental_dilution_non_dim_poisson}, thus we are left with Eqn.s~\ref{eqn:fundamental_dilution_non_dim_eon},~\ref{eqn:fundamental_dilution_non_dim_ion} and~\ref{eqn:electroneutrality_condition}.
We now focus on impedance conditions, i.e. we suppose an off-equilibrium perturbation of the boundary conditions which in turn will slightly affect all unknowns (terms with superscript $(1)$ are much smaller than the terms with superscript $(0)$):
\begin{subequations}
\begin{eqnarray}
\tilde\phi &=& \tilde \phi^{(1)}\\
c_{eon} &=& c_{eon}^{(0)}+c_{eon}^{(1)}  = c_{eon}^{(0)}\left(1+\frac{c_{eon}^{(1)}}{c_{eon}^{(0)}} \right)\\
c_{ion} &=& c_{ion}^{(0)}+c_{ion}^{(1)}= c_{ion}^{(0)}\left(1+\frac{c_{ion}^{(1)}}{c_{ion}^{(0)}} \right)
\end{eqnarray}
\label{eqn:definition_expansion}
\end{subequations}
\noindent We set $n^{(1)} = \frac{c_{eon}^{(1)}}{c_{eon}^{(0)}}$ and $p^{(1)} = \frac{c_{ion}^{(1)}}{c_{ion}^{(0)}}$ and suppose $c_{eon}^{(0)}$, $c_{ion}^{(0)}$ are uniform and $\phi^{(0)}=0$. If we also use the definitions of Eqn.~\ref{eqn:definition_expansion} in the Eqns.~\ref{eqn:fundamental_dilution_non_dim_eon} and \ref{eqn:fundamental_dilution_non_dim_ion},  we obtain:

\begin{subequations}
\begin{eqnarray}
\frac{\tau_n}{\tau}\partial_t \left( 1+n^{(1)} \right)+\nabla_{\tilde x}\cdot \left(-\nabla_{\tilde x}(1+n^{(1)})+(1+n^{(1)})\nabla_{\tilde x} \tilde \phi^{(1)} \right)=0 \\
\frac{\tau_p}{\tau}\partial_t \left(1 +p^{(1)} \right)-\nabla_{\tilde x}\cdot \left( \nabla_{\tilde x}(1+p^{(1)})+2(1+p^{(1)})\nabla_{\tilde x} \tilde \phi^{(1)} \right)=0
\end{eqnarray}
\label{eqn:electroneutral_unexpanded}
\end{subequations}
\noindent If we retain in Eqn.~\ref{eqn:electroneutral_unexpanded} only first order terms, we get:


\begin{subequations}
\begin{eqnarray}
\frac{\tau_n}{\tau}\partial_{\tilde t} n^{(1)} -\triangle_{\tilde x}n^{(1)}+\triangle_{\tilde x}\tilde \phi^{(1)}=0\\
\frac{\tau_p}{\tau}\partial_{\tilde t} p^{(1)}-\triangle_{\tilde x}p^{(1)}-2\triangle_{\tilde x} \tilde \phi^{(1)}=0
\end{eqnarray}
\label{eqn:electroneutral_expanded}
\end{subequations}

The electroneutrality condition, Eqn.~\ref{eqn:electroneutrality_condition}, at first order gives that $\displaystyle p^{(1)} = \frac{1}{2} \frac{c_{eon}^{(0)}}{c_{ion}^{(0)}} n^{(1)} = \frac{1}{2}\frac{\bar n}{\bar p}n^{(1)}$. Thus defining:

\begin{subequations}
\begin{eqnarray}
\tau_n^\star =\frac{\tau_n+\frac{\bar n}{4 \bar p} \tau_p}{1+\frac{\bar n}{4 \bar p}}\\
\tau_\phi^\star =\frac{\tau_p-\tau_n}{1+\frac{4 \bar p}{\bar n}}
\end{eqnarray}
\label{eqn:defn_tau_star}
\end{subequations}
helps rewrite the Eqn.~\ref{eqn:electroneutral_expanded} as:
\begin{subequations}
\begin{eqnarray}
\frac{\tau_n^\star}{\tau} \partial_{\tilde t}n^{(1)}-\triangle_{\tilde x} n^{(1)}=0\\
\frac{\tau_\phi^\star}{\tau} \partial_{\tilde t}n^{(1)}-\triangle_{\tilde x} \tilde \phi^{(1)}=0
\end{eqnarray}
\label{eqn:to_solve}
\end{subequations}

\subsection{Boundary Conditions}
It follows from symmetry, Fig.~\ref{fig:mat_domain}, that  $\partial_{\tilde x} \tilde\phi^{(1)} =\partial_{\tilde x} \tilde n^{(1)}=0$ on $\Gamma_2$ and $\Gamma_3$. Since the metal is ion-blocking, $\frac{1}{2}\frac{\bar n}{\bar p}\partial_{\tilde y}n^{(1)}+2\partial_{\tilde y} \tilde \phi^{(1)}=0$ will be satisfied on $\Gamma_4$. We assume as well that the response of the metal to an electric perturbation is fast compared to the MIEC, from this it follows that we can take the electric potential $\tilde \phi^{(1)}$ uniform on $\Gamma_4$. Thank to linearity and given the impedance setting, we can choose $\tilde \phi^{(1)}=\frac{1}{\sqrt{2 \pi}}\Re\left(e^{i\omega\tau\tilde t}\right)$ on $ \Gamma_4$  and  $\tilde \phi^{(1)}=n^{(1)}= 0$ on $ \Gamma_1$.

We assume the chemistry due to the reactions on $\Gamma_5$ has a finite speed and that it is correctly characterized by a one-step reaction \cite{Ciucci-09}. For simplicity we start from:
\begin{equation}
H_2({\rm gas}) \rightleftharpoons H_2O({\rm gas}) + V_O^{\bullet\bullet}+2e'
\label{global_reaction}
\end{equation}

We also remark \cite{Ciucci-09} that the rates of injection of vacancies $\dot \omega_{ion, S}$ and electrons $\dot \omega_{eon, S}$ at $\Gamma_5$ satisfy (subscript $S$ indicates surface) the following two equations:

\begin{equation}
\begin{array}{lcl}
\displaystyle\dot \omega_{ion, S} &=& \frac{1}{2}\dot\omega_{eon, S} \\
\displaystyle\dot\omega_{ion,S} &=& k_f \tilde p_{H_2} - k_r \tilde p_{H_2O} c_{ion}c_{eon}^2
\end{array}
\end{equation}
\noindent where $k_f$ is the forward rate of the reaction in Eqn.~\ref{global_reaction} and $k_r$ is the reverse rate.

%
%



The latter gives, under small perturbation assumptions \cite{Ciucci-09}, a Chang-Jaff\'e boundary condition \cite{ISI:A1952UC10100004}:
\begin{equation}
-\dot\omega_{eon, S}^{(1)} =  4\frac{D_{ion}}{l_c} \tilde k_f^0\tilde p_{O_2}^{1/4} \left(1+\frac{c_{eon}^{(0)}}{4 c_{ion}^{(0)}}\right) \tilde p_{H_2}n^{(1)}
\label{eqn:chang_jaffe}
\end{equation}

We suppose $k_f = 2\frac{D_{ion}}{l_c}\tilde k_f$ and $\tilde k_f = \tilde k_f^0\tilde p_{O_2}^{\beta}\times\frac{\# \text{particles}}{m^3} $, 
\footnote{
The order of magnitude of $k_f$ is given by:
$p_{O_2}= 10^{-24}$, $l_c = 10^{-5} m$, $D_{ion} = 10^{-10} m^2/s$ and $\tilde k_f^{0} \approx 10^{32}$, so $k_f\approx10^{32}\times \frac{10^{-10}}{10^{-5}}\times10^{-6}$$=$$10^{21} \frac{\# \text{particles}}{m^2}\approx10^{-3} \frac{mol}{m^2}$$\approx 10^{-7}\frac{mol}{cm^2}$
} where we choose $\beta = 1/4$ \cite{Ciucci-09}.

Hence the $y$-flux of electrons and vacancies satisfies the following expression along $\Gamma_5$: $\displaystyle{\bf j}_{eon}^P\cdot {\bf e}_y = 2 \displaystyle{\bf j}_{ion}^P\cdot {\bf e}_y = -  \dot\omega_{eon,S}$. If we define $\displaystyle \tilde A_\phi =\tilde k_f \frac{\tilde p_{H_2}}{c_{ion}^{(0)}}\left( 1-\frac{D_{ion}}{D_{eon}}\right )$ and $\displaystyle \tilde A_n = \tilde k_f \frac{\tilde p_{H_2} }{c_{ion}^{(0)}}\left( 1+ 4 \frac{D_{ion} c_{ion}^{(0)}}{D_{eon} c_{eon}^{(0)}}\right)$, we can rewrite the boundary conditions on  on $\Gamma_5$ as $\partial_{\tilde y} \tilde \phi^{(1)} = \tilde A_\phi n^{(1)}$ and  $\partial_{\tilde y} n^{(1)} =\tilde A_n n^{(1)}$.

\subsection{Weak Formulation of the Model}

If we Fourier transform Eqn.s~\ref{eqn:to_solve} and the boundary conditions with respect to $\tilde t$~\footnote{We choose unitary Fourier transform $\hat f (\omega) = \frac{1}{\sqrt{2\pi}}\int_{-\infty}^{\infty} f(x) e^{-i\omega x}\ud x$}, we find the following system of equations ($\hat{(\cdot)}$ indicates Fourier transformed quantity)~\footnote{We factored out the Dirac distribution that comes out of Fourier transformation of an exponential} which we call IS equations:
\begin{subequations}
\begin{eqnarray}
i \omega \tau_n^\star \hat n^{(1)}-\triangle \hat n^{(1)}=0\\
i \omega \tau_\phi^\star \hat n^{(1)}-\triangle\hat \phi^{(1)}=0
\end{eqnarray}
\label{eqn:to_solve_FT}
\end{subequations} 
with boundary conditions:
\begin{equation}
\left \{  \begin{array}{lclcl}
\hat \phi^{(1)} = 0 &{\wedge}& \hat n^{(1)} = 0& {\rm on } & \Gamma_1 \\
\partial_{\tilde x} \hat \phi^{(1)} =0& {\wedge}&\partial_{\tilde x} \hat n^{(1)} =0  & {\rm on }& \Gamma_2 \quad {\wedge } \quad\Gamma_3 \\
\hat\phi^{(1)}= 1& {\wedge} &\partial_{\tilde y} \hat n^{(1)}=-4 \frac{\bar p}{\bar n}\partial_{\tilde y} \hat \phi^{(1)} & {\rm on } & \Gamma_4 \\
\partial_{\tilde y} \hat \phi^{(1)} = \tilde A_\phi \hat  n^{(1)}& {\wedge} &\partial_{\tilde y} \hat n^{(1)} =\tilde A_n n^{(1)}&{\rm on } &  \Gamma_5
\end{array}
\right.
\label{eqn:BC_to_solve_FT}
\end{equation}

We can recast the Eqn.~\ref{eqn:to_solve_FT} and \ref{eqn:BC_to_solve_FT} in weak form taking as test functions $m_{Re}$, $m_{Im}\in H^1(\Omega\setminus\Gamma_1)$, $\psi_{Re}$, $\psi_{Im} \in H^1(\Omega\setminus(\Gamma_1\cup \Gamma_4))$ \cite{adams-03}:

\begin{subequations}
\begin{eqnarray}
\begin{array}{lcl}
\displaystyle\omega \tau^\star_n \int_\Omega \hat n^{(1)}_{Im} m_{Re} \ud \tilde A& -& \displaystyle\int_\Omega \nabla \hat n^{(1)}_{Re} \cdot \nabla m_{Re} \ud \tilde A+\int_{\Gamma_5} \tilde A_n \hat n^{(1)}_{Re} m_{Re} \ud \tilde A+ \ldots\\
 &-& \displaystyle 4 \frac{\bar p}{\bar n}\int_{\Gamma_4} \partial_{\tilde y} \hat \phi ^{(1)}_{Re} m_{Re} \ud \tilde x =0
\end{array}\\
\begin{array}{lcl}
\displaystyle\omega \tau^\star_n \int_\Omega \hat n^{(1)}_{Re} m_{Im} \ud \tilde A &+& \displaystyle \int_\Omega \nabla \hat n^{(1)}_{Im} \cdot \nabla m_{Im} \ud \tilde A-\int_{\Gamma_5} \tilde A_n \hat n^{(1)}_{Im} m_{Im}  \ud \tilde x \\
&+& 4 \displaystyle \frac{\bar p}{\bar n}\int_{\Gamma_4} \partial_{\tilde y} \hat \phi ^{(1)}_{Im}m_{Im} \ud \tilde x =0
\end{array}\\
\omega \tau^\star_\phi \int_\Omega \hat n^{(1)}_{Im} \psi_{Re} \ud \tilde A - \int_\Omega \nabla \hat \phi^{(1)}_{Re} \cdot \nabla \psi_{Re} \ud \tilde A+\int_{\Gamma_5} \tilde A_\phi \hat n^{(1)}_{Re} \psi_{Re} \ud \tilde x=0 \label{eqn:variation_phi_Re}\\
\omega \tau^\star_\phi \int_\Omega \hat n^{(1)}_{Re} \psi_{Im} \ud \tilde A + \int_\Omega \nabla \hat \phi^{(1)}_{Im} \cdot \nabla \psi_{Im} \ud \tilde A-\int_{\Gamma_5} \tilde A_\phi \hat n^{(1)}_{Im} \psi_{Im}\ud \tilde x=0 \label{eqn:variation_phi_Im}
\end{eqnarray}
\label{eqn:to_solve_FT_weak}
\end{subequations}

with the condition that:
\begin{subequations}
\begin{align}
\hat \phi^{(1)}_{Re} &= 0 &\quad &{\wedge} &\hat \phi^{(1)}_{Im}& = 0 &\quad &{\rm on }  &\Gamma_1 \\
\hat n^{(1)}_{Re} &= 0 &\quad &{\wedge} &\hat n^{(1)}_{Im} &= 0 &\quad&{\rm on }  &\Gamma_1 \\
\hat \phi^{(1)}_{Re} &= 1 &\quad &{\wedge} &\hat \phi^{(1)}_{Im}& = 0 &\quad &{\rm on }  &\Gamma_4
\end{align}
\label{eqn:BC_to_solve_FT_FEM}
\end{subequations}
\noindent It is easy to show that the sum of the Eqn.s~\ref{eqn:to_solve_FT_weak} is bounded and thus the bilinear form associated to the weak formulation of Eqn.s~\ref{eqn:to_solve_FT} with \ref{eqn:BC_to_solve_FT} is continuous. Further, the problem is weakly-coercive hence it admits one unique solution \cite{Agmon-65}.

\subsection{Numerical Solution Procedure for the 2D Case}
In order to solve numerically the Eqn.s~\ref{eqn:to_solve_FT_weak} with boundary conditions  Eqn.s~\ref{eqn:BC_to_solve_FT_FEM} we employ an h-adapted finite element method (FEM), implemented with FreeFem++ \cite{hec-pir-2007}. The governing equations are discretized on a triangular unstructured mesh using quadratic continuous basis functions with a centered third order bubble. We use a direct method to solve the linear system following integration of Eqn.s~\ref{eqn:to_solve_FT_weak} in the discretized mesh. Then the mesh is adaptively refined nine times for each case. The a posteriori adaptation is performed the first six times against the 4 dimensional vector $\left(\nabla \Re\left[ \hat \mu_{eon}^{(1)}\right], \nabla \Re\left[ \hat \mu_{ion}^{(1)}\right]\right)$ and subsequently against $\eta_\varepsilon$, see Appendix \ref{appendix_A}. The h-adaptation ensures high regularity of the $H^1$ a posteriori estimator \cite{Brenner-00}, locally below $10^{-5}$,  and it guarantees that the mesh is finer where sharper gradients occur. Independently of frequency, mesh adaptivity results in coarseness everywhere except in the vicinity of the interfaces, in particular the refinement increases towards the triple-phase boundary (the intersection of metal, oxide and gas phases, which is though to be a particularly active site for electrochemical reactions \cite{Trovarelli-02} \cite{Mogensen200063}); this fact indicates strong non-linearities around that area.
Finally we note that FreeFem++ execution time is comparable to custom-written C++ code and its speed is enhanced by the utilization of fast sparse linear solvers such as the multi-frontal package UMFPACK \cite{992206}.  Due to the sparsity of the problem we make extensive use of this last feature. 

We further note that the utilization of asymptotic expansion and Fourier transformation techniques, while guaranteeing linearity, has a great speed advantage over direct sinusoidal \cite{Goodwin-05} and step relaxation techniques \cite{bessler:B1186}. Further, this method can be directly used to examine chemical reactions within the cell and draw directly conclusions about fast and rate-limiting chemical reactions. Also, this procedure lends itself to direct error estimation and its implementation can be done automatically for a time-dependent problem \cite{Ciucci-07}.

\subsection{1D case: Analytical Solution}
Since we also aim at comparing the 1D and 2D solutions, it is beneficial to revisit the 1D solution of Eqn.s~\ref{eqn:to_solve_FT} \cite{macdonald:4982}. The solution $(\hat n^{(1)}, \hat \phi^{(1)})$ will satisfy (if $\omega \neq 0$):

\begin{subequations}
\begin{eqnarray}
\hat n^{(1)} = \sum_{\pm} a_{\pm} e^{\pm \sqrt{i} \sqrt{\tau_n^\star \omega} \tilde y}\\
\hat \phi^{(1)} = \hat \phi^{(1)}_0  + (\hat \phi^{(1)}_0)' \tilde y+\frac{\tau_\phi^\star}{\tau_n^\star} \hat n^{(1)}
\end{eqnarray}
\label{eqn:soln_1D_most_general}
\end{subequations}
\noindent where for simplicity we indicate $\sqrt{i}= e^{i\frac{\pi}{4}}$. The boundary conditions, as in the 2D case, at $\tilde y = 0$ ($\Gamma_1$) are:
\begin{equation}
\hat \phi^{(1)}= 0  \quad \wedge \quad \hat n^{(1)}= 0
\end{equation}
\noindent The latter can help rewrite Eqn.s~\ref{eqn:soln_1D_most_general} as:
\begin{subequations}
\begin{eqnarray}
\hat n^{(1)} = 2a_+\sinh\left(\sqrt{i} \sqrt{\tau_n^\star \omega} \tilde y\right)\\
\hat \phi^{(1)} =  (\hat \phi^{(1)}_0)' \tilde y+2a_+\frac{\tau_\phi^\star}{\tau_n^\star} \sinh\left(\sqrt{i} \sqrt{\tau_n^\star \omega} \tilde y\right)
\end{eqnarray}
\label{eqn:soln_1D_with_symm}
\end{subequations}


\noindent If we set $\gamma_\phi =\frac {R_{ion}^\perp e l_c D_e c_{eon}^{(0)}} {U_T\left( 1+\frac{1}{4}\frac{\bar n}{\bar p}\right)}$ and $\gamma_n =\frac{1}{4}\frac{\bar n}{\bar p} \gamma_\phi$, then at  $\tilde y =l_2$ we have the following conditions \cite{ISI:000232773100001}:

\begin{equation}
\displaystyle \hat \phi^{(1)}= 1  \quad \wedge \quad \displaystyle\hat n^{(1)}+\gamma_\phi \frac{d\hat \phi^{(1)}}{d\tilde y}+ \gamma_n \frac{d\hat n^{(1)}}{d\tilde y}=0 
\label{eqn:1D_BC_after_symm}
\end{equation}

The boundary conditions Eqn.~\ref{eqn:1D_BC_after_symm} will lead to the determination of $a_+$ and  $(\hat \phi^{(1)}_0)' $ in Eqn.~\ref{eqn:soln_1D_with_symm} and the 1D model leads to impedance of the form \cite{ISI:000252830600012} \cite{wei_lai_thesis} \cite{Macdonald-05}:

\begin{equation}
Z_{1D}(\omega,\tilde{p}_{O_{2}},T)=R_{\infty}+\left(R_{0}-R_{\infty}\right)\left(1+\frac{R_{ion}+R_{eon}}{2R_{ion}}\right)\frac{\tanh s}{s+ \displaystyle\frac{R_{ion}+R_{eon}}{2R_{ion}^\perp}\tanh s}
\label{eq:Z-1d}
\end{equation}
\noindent where all the relevant terms are reported in Table~\ref{tab:definitions_1D}.

\section{Results}
\subsection{Comparison to Experiments}
The electron electrochemical potential drop across the sample, i.e. the electron electrochemical potential difference between the top and bottom electrodes ($\Gamma_4$ and its symmetric reflection), is given by the following expression:

\begin{equation}
\hat V^{(1)} = 2 U_T \left[<\left(\hat \mu_e^{(1)}\right)^\star>_{\Gamma_4}-<\left(\hat \mu_e^{(1)}\right)^\star>_{\Gamma_1} \right]
\end{equation}
\noindent where $<a>_\Lambda$ indicates the average of the quantity $a$ over the set $\Lambda$.
At first order the $\star$-electrochemical potential is given by $\left(\hat \mu_e^{(1)}\right)^\star =\hat \phi^{(1)}-\hat n^{(1)}$. The electric current density at the the two ends of the circuit is:
\begin{align}
\hat j^{(1)} &= \frac{D_{eon} e c_{eon}^{(0)} \int_{\Gamma_4}\nabla_{\tilde x}\frac{\tilde \mu_{eon}^{(1)}} {k_b T}\cdot \mathbf e_y \ud \tilde x}{\left(W_1+W_2 \right)l_c }
\end{align}

\noindent  Hence, the 2D impedance is given by the expression:
\begin{equation}
Z_{2D}(\omega,\tilde{p}_{O_{2}},T)=\displaystyle{\hat V^{(1)}}/{\hat j^{(1)}}\label{eq:Z-2d}
\end{equation}

\begin{figure}
\scalebox{0.8}{\includegraphics{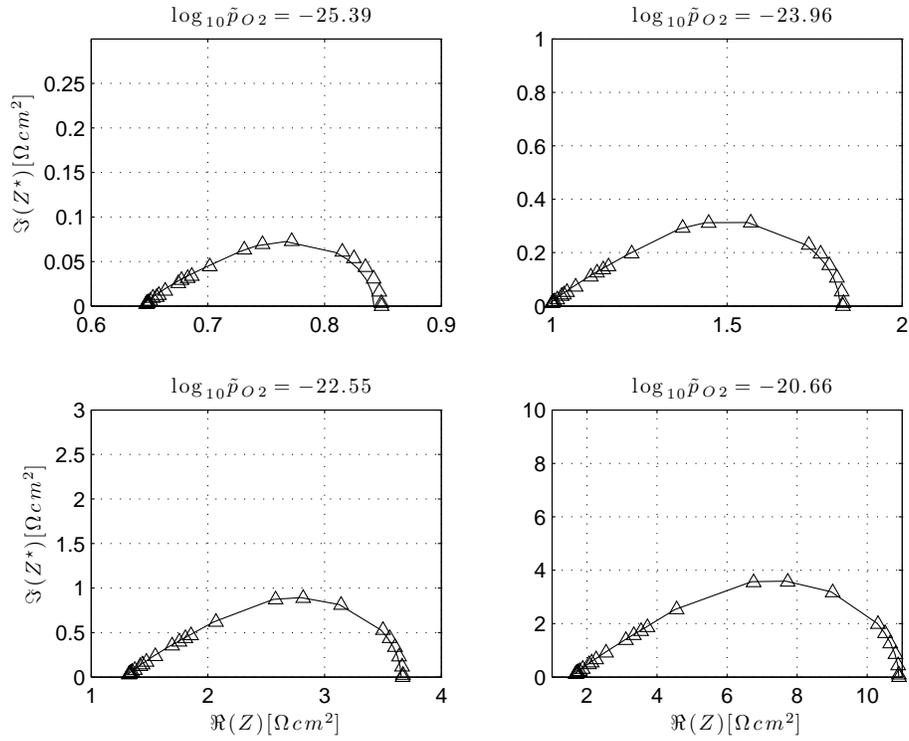}}
\caption{\label{fig:impedance_fittings} The triangle indicated fitted computations while the solid line is the experimental value. The results are presented at $650^oC$ varying the $\tilde p_{O_2}$ partial pressure.}
\end{figure}

\noindent We define the error of the 2D impedance $Z_{2D}$ with respect to experimental impedance $Z_{1D}$ spectra Eqn.~\ref{eq:Z-1d} as follows: 
\begin{equation}
\varepsilon_F(\omega,\hat{p}_{O_{2}},T)=\left|1-\frac{Z_{2D}(\omega,\tilde{p}_{O_{2}},T)}{Z_{1D}(\omega,\tilde{p}_{O_{2}},T)}\right|
\end{equation}
\noindent For every data point, uniquely defined by the couple $(\tilde p_{O_2}, T)$, we fit the 2D data against the measured 1D equivalent circuit data in \cite{ISI:000232773100001} by minimizing $\varepsilon_F(\omega,\tilde{p}_{O_{2}},T)$ with respect to the surface reaction constant $\tilde{k}_{f}^{0}=A\tilde{p}_{O_{2}}^{\alpha}$, which is a function of both$O_2$ partial pressure and temperature.  We remark that $\tilde{k}_{f}^{0}$ is the sole parameter we allow to vary in this procedure and all other data is obtained from the literature and presented in Tab.~\ref{tab:fitted_kf0}. With only one parameter variation, we obtained excellent agreement between experimental results and 2D calculations, i.e. $\varepsilon_F(\omega,\hat{p}_{O_{2}},T)<2\%$. As an example, 2D results at four different oxygen partial pressures and at 650$^{\circ}$C are shown in Fig.~\ref{fig:impedance_fittings}.
We computed the $\tilde{k}_{f}^{0}$ by minimizing the $\varepsilon_F$ for a total of 28 cases (7 pressures times 4 temperature). We report in Tab.~\ref{tab:fitted_kf0} the results of linear regression of these minimizing values (each line is derived on keeping the temperature fixed and varying $\tilde p_{O_2}$). We also write in Tab.~\ref{tab:fitted_kf0}, the 95\% confidence intervals for  the fitting of $A$, i.e., $A\approx\bar{A}\pm\varepsilon_A$, and $\alpha$, i.e., $\alpha=\bar{\alpha}\pm\varepsilon_\alpha$; we finally report the root mean square error $\sigma$ and the adjusted $R$-squared \cite{Draper-98}, regarding the latter, a value close to unity indicates a perfect fit while negative values indicate poor data correlation. Directly from analysis of Tab.~\ref{tab:fitted_kf0} we deduce that $\tilde k_f^{0}$ fitting to a straight line is reasonable for   "high" temperatures ($T\geq 550^oC$). We note that $\tilde k_f^0$ is temperature-dependent via $\bar{A}$ ($\bar A$ decreases with $T$). Furthermore $\tilde k_f^{0}$ is slightly pressure dependent via the coefficient $\alpha$, the average value of $\bar{\alpha}\approx0.05\ge0$, however the error is of the same order of the slope. Hence the total rate of reaction is very likely to be $\dot \omega_{eon, S} \propto \tilde p_{O_2}^{-1/4+\beta}$ where $\beta$ is somewhere in the set $[0,0.1]$, most likely equal to $0.05$.

\subsection{The Polarization Resistance in Frequency Space}
One of the goals of fuel cell science is to understand and possibly reduce the polarization resistance, i.e. that portion of the resistance due to electric field effects at interfaces. For that purpose it is key to identify and understand the main processes that intervene in the definition of this quantity. Specifically, the area specific polarization resistance for our system is defined as \cite{Ciucci-09}:
\begin{equation}
Z_{ion}^\perp = U_T \frac{<\hat \mu_{ion}^\star>_{\Gamma_5}- <\hat \mu_{eon}^\star>_{\Gamma_4}}{\hat j_{IP}^{(1)}} \label{eq:R_ion_perp}
\end{equation}

\noindent where $\hat j_{IP}^{(1)} =\frac{1}{W_1+W_2} \int_{\Gamma_5} \dot\omega_{eon, S} \ud x$ is the ionic contribution to the area specific current. The $Z^\perp_{ion}$ can be understood as the sum of a surface $Z_{surf}$ and a bulk polarization resistance, $Z_{bulk}= Z_{ion}^\perp-Z_{surf}$, where the $Z_{surf}$ is the portion of the area-specific resistance due to effects of the exposed boundary $\Gamma_{5}$ and it is given by:
\begin{equation}
Z_{surf} = U_T \frac{<\hat \mu_{ion}^\star>_{\Gamma_5}- <\hat \mu_{eon}^\star>_{\Gamma_5}}{\hat j_{CP}^{(1)}} \label{eq:Z_surf} 
\end{equation}

\noindent In our model, by definition, the $Z_{surf}\in \mathbb R^+$ is proportional to $\left(1+W_1/W_2\right)$ and inversely proportional to both $\tilde p_{H_2}$ and $k_f$:

\begin{equation}
Z_{surf} =\frac{1}{2} \left(1+\frac{W_1}{W_2}\right)\frac{U_T}{e k_f \tilde p_{H_2}} \label{eq:Z_surf_explicit} 
\end{equation}

\begin{figure}
\scalebox{0.6}{\includegraphics{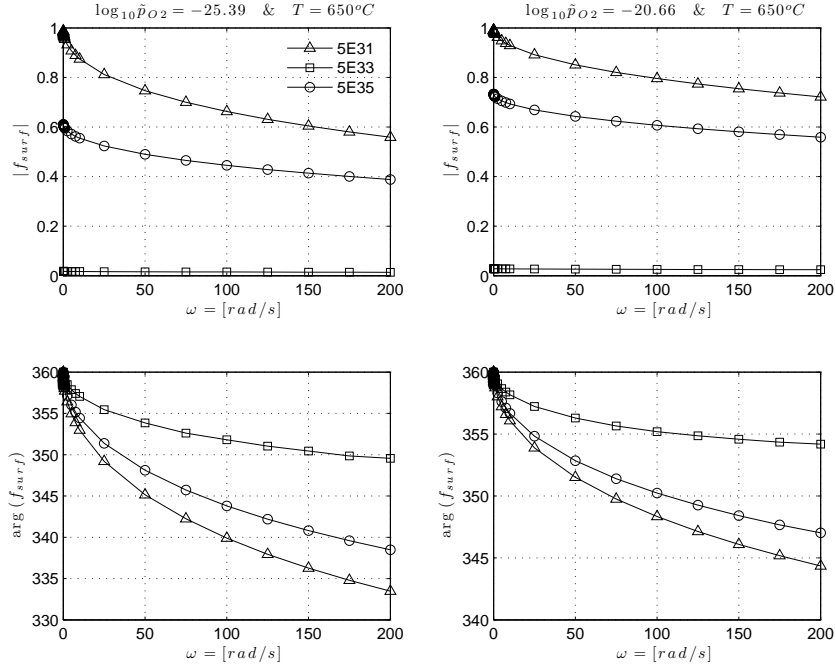}}
\caption{\label{fig:f_surf_plot} Plot of $f_{surf} = \frac{R_{surf}}{R_{ion}^\perp}$ as a complex function of $\omega$. We present two cases, both at $650^oC$, the one to the left at very reducing conditions $\tilde p_{O_2} = 10^{-25.32}$ and the one to the right at $\tilde p_{O_2} = 10^{-20.66}$, parametrized versus $\tilde k_f^{(0)}$.}
\end{figure}

The fraction $f_{surf} = \displaystyle \frac{Z_{surf}}{Z_{ion}^{\perp}}$ indicates what portion of the polarization impedance is due to surface effects. From Fig.~\ref{fig:f_surf_plot} we note two fundamental facts: first, as we expect, at "lower" injection rates the $f_{surf}$ increases, physically this means that that if the chemistry is sufficiently slow it will dominate the polarization resistance leading to an $f_{surf}$ of approximately unity. 
Second, we notice frequency dependent behavior of $R^\perp_{ion}$. Our computations show that $f_{\mathrm{surf}}$ decreases with $\omega$, while the dephasing between $Z_{\mathrm{surf}}$ and $Z_{\mathrm{ion}}^{\perp}$, described by $\arg(f_{\mathrm{surf}})$, increases with $\tilde{k}_{f}^{0}$ and decreases with $\omega$. The behavior of $f_{\mathrm{surf}}$ in phase space clearly shows that $Z_{surf}$ includes two interrelated processes: 
\begin{enumerate}
\item{reactions on the surface exposed to the gas;}
\item{transport  of charged species in MIEC.}
\end{enumerate}  
Within this framework, as $\omega$ increases, the losses in the polarization due to drift diffusion increase and surpass the (constant) reaction or surface losses.

\subsection{Analysis of the 2D Solution}
\subsubsection{Qualitative Considerations}
\begin{figure}
\scalebox{0.4}{\includegraphics{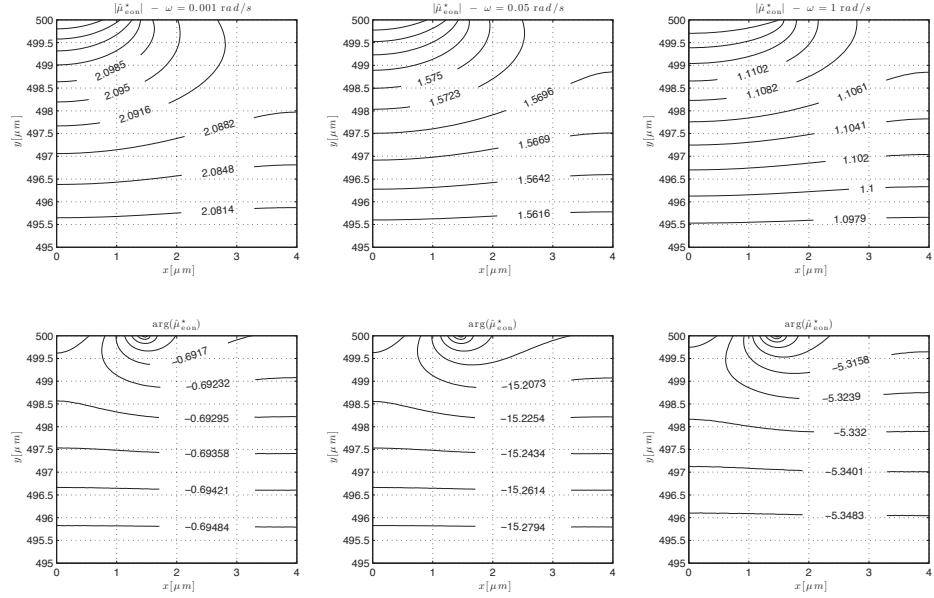}}
\caption{\label{fig:var_omega_mue} Plots of the complex electrochemical potential of electrons $\hat \mu_{eon}(x,y,\omega)$ as a function of $x$ and $y$ in the case where $T = 650^oC$ and $\tilde p_{O_2} = 10^{-25}$. In the top panels we depict its absolute value $|\hat \mu_{eon}|$ while at the bottom we show its argument $\arg(\hat \mu_{eon})$. The applied frequency is increased from left to right, going from 0.001 rad/s to 1 rad/s.}
\end{figure}

\begin{figure}
\scalebox{0.4}{\includegraphics{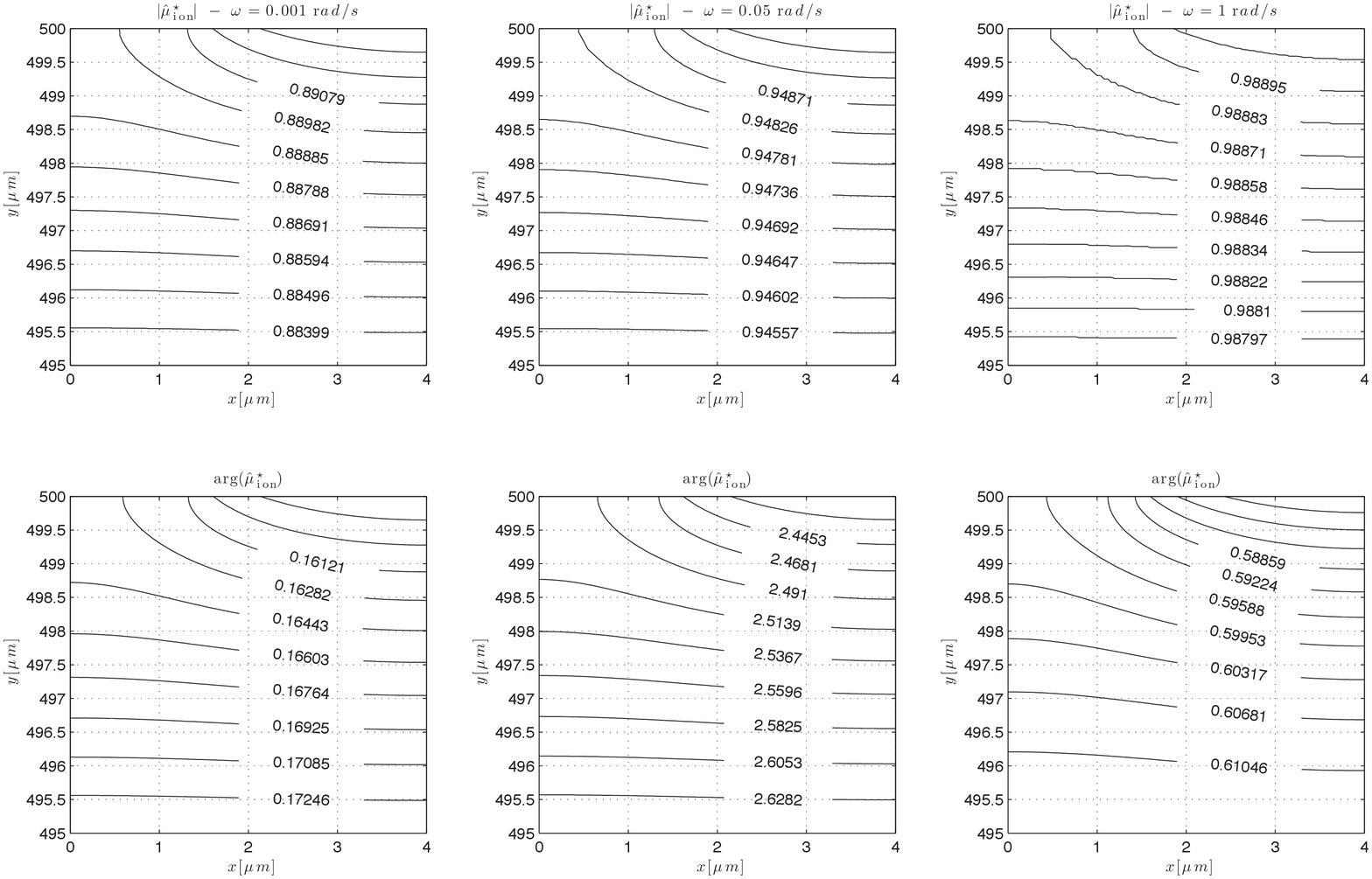}}
\caption{\label{fig:var_omega_muv} Similar to Fig.~\ref{fig:var_omega_mue}, we depict the complex electrochemical potential of ions $\hat \mu_{ion}(x,y, \omega)$ where at the top we show $|\hat \mu_{ion}|$ and at the bottom $\arg(\hat \mu_{ion})$. The conditions are the same as Fig.~\ref{fig:var_omega_mue} and so is the frequency range.}
\end{figure}

We can then use the framework to study the two complex electrochemical potentials $\hat{\mu}_{\mathrm{eon}} =\hat n^{(1)}-\hat \phi^{(1)}$ and $\hat{\mu}_{\mathrm{ion}}=\hat \phi^{(1)}+\frac{\bar n}{2 \bar p}\hat n^{(1)}$ as functions of frequency. In Figs.~\ref{fig:var_omega_mue} and \ref{fig:var_omega_muv} we plot the 2D distributions of the latter in the computational domain at $T=650^{\circ}$C, $\tilde{p}_{O_{2}}=10^{-25}$ and $\tilde{k}_{f}^{0}=10^{32}$ with frequency $\omega$ increasing from $10^{-3}$ to $10^5~rad/s$ . Thank to the Figs.~\ref{fig:var_omega_mue} and \ref{fig:var_omega_muv}, we can address the qualitative behavior of the solution. We first analyze the qualitative distribution of fluxes: from the gradient of $|\hat \mu_{eon}|$, which gives an idea of electron flux, that electrons flow from the gas$|$ceria interface $\Gamma_5$ onto the ceria$|$metal interface $\Gamma_4$ through a cross-plane current $\displaystyle \hat I_g^{CP}$, and concurrently electrons flow onto the ceria$|$metal interface $\Gamma_5$ from its mirror symmetric counterpart. Similarly the MIEC$|$metal interface is blocking to vacancies, hereby the vacancies correctly flow from the bottom to the top ceria$|$gas interface $\Gamma_5$.
It is also clear that the complex potential of the electrons $\hat{\mu}_{\mathrm{eon}}$ changes significantly as $\omega$ increases, while $\hat{\mu}_{\mathrm{ion}}$ is relatively unaffected.
The penetration depth, which is defined as the vertical displacement from $\Gamma_{4}$ where surface electrons can penetrate into the bulk, decreases with $\omega$ as the 1D model hints (in Eqn.s \ref{eqn:soln_1D_with_symm} the solution decays exponentially with $1/\sqrt{\tau_n^\star \omega}$). As $\omega$ increases, the dephasing of $\hat{\mu}_{\mathrm{eon}}$ first increases and then decreases and it is weakly dependent upon the distance from $\Gamma_4$, or conversely, the penetration depth into the MIEC. We notice that the same dephasing increases and then decreases for $\hat \mu_{ion}$.
However, while for the vacancies, the behavior of $|\hat \mu_{ion}|$ and  $\arg(\hat \mu_{ion})$ is qualitatively the same, this is not the case for the electrons, where through a wide array of $\omega$'s, the qualitative behavior of $|\hat \mu_{eon}|$ and $\arg(\hat \mu_{eon})$ is distinctly different. 

Deriving the electronic and ionic currents from the computations requires some care and it will not simply be $\nabla |\hat \mu_m|$
For example, for electrons, we note that:
\begin{equation}
\tilde \mu_{eon}^{(1)} = \left(n^{(1)}-\phi^{(1)}\right)e^{i\omega t}
\end{equation}
We will call the complex current $\mathbf j^{\mathbb C}_{eon}$:
\begin{equation}
\mathbf j^{\mathbb C}_{eon} =c_{eon}^{(0)}D_{eon} \mathcal F ^{-1}[\nabla \hat \mu_{eon}^{(1)}] 
\end{equation}
the physical current will be \footnote{We remark that for complex valued function $\mu$ in general we have $\text{abs}(\nabla\tilde \mu) \neq \nabla\left(\text{abs}\left(\mu\right)\right)$}:
\begin{eqnarray}
\mathbf j_{eon} &=&\Re \left(\mathbf j^{\mathbb C}_{eon}\right) 
\end{eqnarray}

\begin{figure}
\scalebox{0.6}{\includegraphics{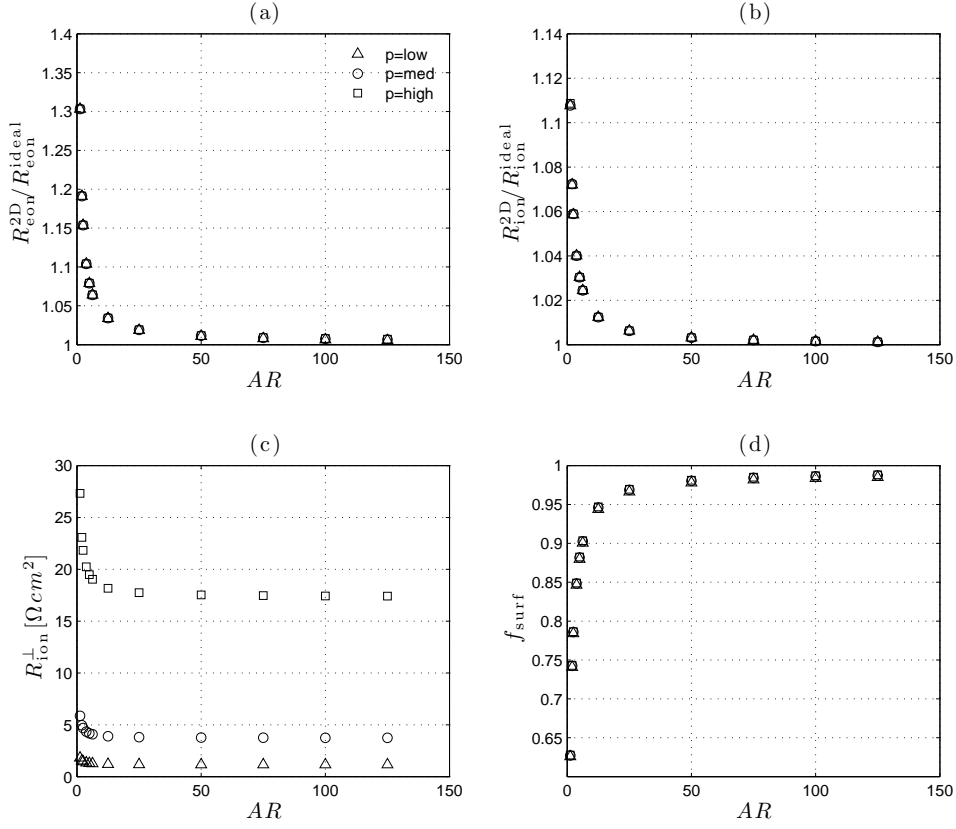}}
\caption{\label{fig:two_by_two} Deviation of the 2D model from 1D behavior as a function of the aspect ratio $AR = \left( W_1+W_2\right)/l_2$. We consider the case where $\tilde k_f^{(0)}=10^{32}$, $T = 650^oC$ and we set $\tilde p_{O_2} =10^{-25.32}$(p=low), $\tilde p_{O_2} =10^{-23.34}$ (p=med), $\tilde p_{O_2} =10^{-20.66}$. The $R_{eon}^{2D}$ and the $R_{ion}^{2D}$ monotonically approach their 1D (ideal) value if $AR$ is sufficiently large. $R_{ion}^\perp$ increases with decreasing the $AR$ while the $f_{surf}$ decreases, which indicates that if the thickness is reduced enough, the $R_{ion}^\perp$ is not just surface dominated.}
\end{figure}

In order to compare the 1D and 2D solutions qualitatively, we first focus on the case $\omega=0$ where $\tilde k_f^{(0)}=10^{32}$, and we shrink the size of the slab while keeping the same framework and model parameters. This corresponds to a decrease of the aspect ratio of the sample defined as $AR= \frac{l_2}{W_1+W_2}$. We show in Fig.~\ref{fig:two_by_two} the results of the computations in the case where the conditions are very reducing. We depict what happens to $R_{ion}$, $R_{eon}$, $R_{ion}^\perp$ and $f_{surf}$ as $AR$ changes. We notice that decreasing $AR$ corresponds to an increase in effective electronic and ionic resistance compared to the ideal case computed according to Tab.~\ref{tab:definitions_1D} which in turn corresponds to $AR \rightarrow \infty$. Deviations from ideality occur already for $AR\approx 25$, hence even for reasonably large $AR$ the ionic and electronic resistances deviate from the ideal 1D case, this is clearly shown in Fig.s~\ref{fig:two_by_two} a and b. The same applies to the polarization resistance $R_{ion}^\perp$, Fig.~\ref{fig:two_by_two}c, which is flat above $AR \approx 25$, below this value $R_{ion}^\perp$ sharply increases due to bulk polarization effects. As the deviation from the 1D setting starts, not only ionic and electronic resistivities change, but so does the relative importance of surface and drift diffusion effects. Hence the polarization resistance is thickness-dependent, and the dependence is due to the emergence of two-dimensional effects. The increase in drift diffusion resistance due to the motion of electrons from $\Gamma_5$ to $\Gamma_4$ is also shown in the $f_{surf}$ which increases with the $AR$ reaching unity for $AR\rightarrow \infty$.
This effect is even clearer if we plot the electrochemical potentials of electrons and vacancies at $\omega =0$, we note a shrinking of the affected area as the sample thickness decreases corresponding to an increase of polarization resistance. This effect is purely 2D and cannot be studied using a 1D model.

\subsubsection{Quantitative Analysis}
In order to compare the 1D and 2D solution quantitatively we define the following two functionals:
\begin{subequations}
\begin{eqnarray}
\nu\left[\hat \mu_{1D}, \hat\mu_{2D}, \tilde y, \omega\right] &=&\frac{1}{W_1+W_2}  \frac{\displaystyle \int_{y'=\tilde y} \left| \hat\mu_{1D}(y', \omega)-\hat\mu_{2D}(\tilde x, y',  \omega)\right| \ud \tilde x}{\left| \hat\mu_{1D}(l_2, \omega)\right|}  \\
\zeta\left[ \hat\mu_{1D}, \hat\mu_{2D}, \tilde y, \omega\right] &=&\frac{1}{W_1+W_2} \frac{\displaystyle \left|\int_{y'=\tilde y}\left(\hat\mu_{1D}(y', \omega)-\hat\mu_{2D}(\tilde x, y', \omega)\right) \ud \tilde x\right|}{\left| \hat\mu_{1D}(l_2, \omega)\right|} 
\label{eqn:defn_functionals}
\end{eqnarray}
\end{subequations}

\noindent The functional $\nu$ describes the "pointwise" distance between 1D and 2D solutions of $\hat \mu$ at a section $\tilde y$ and the functional $\zeta$ describes the "average" distance between 1D and 2D descriptions. Physically $\nu$ indicates how far apart the 1D and 2D electrochemical potential are, while $\zeta$ "measures" the soundness of fitting a 1D case with the 2D model. We can examine the applicability of the 1D approximation for data fitting via $\zeta$.

\begin{figure}
\scalebox{0.4}{\includegraphics{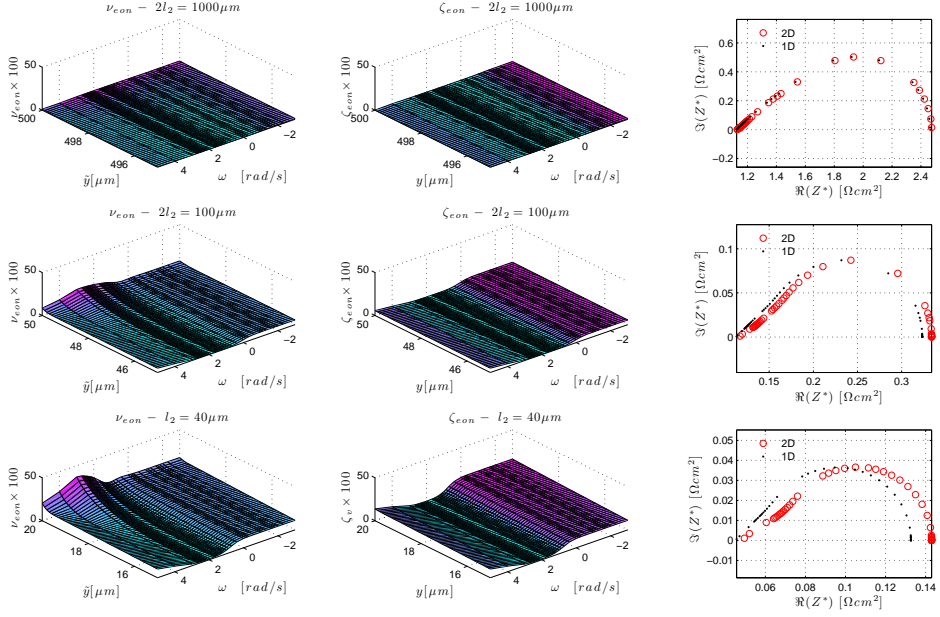}}
\caption{\label{fig:var_AR} Plots of the $\nu$ and $\zeta$'s of the electrochemical potential of electrons (plots are shown up to $5 \mu m$ from $\Gamma_4$ and $\Gamma_5$) as function of $y$ and $\omega$ and of the impedance spectra as  the aspect ratio changes (each line corresponds to a different aspect ratio, $2l_2=1000\mu m$, $2l_2=100\mu m$ and $2l_2 = 40 \mu m$ correspond respectively to $AR = 125$, $AR = 12.5$ and $AR = 5$).  A decrease of the aspect ratio corresponds to an increase of both $\nu$ and $\zeta$ and an increase between the (ideal) 1D impedance and the 2D impedance spectra.}
\end{figure}

In order to further compare the 2D model and 1D model and demonstrate the importance of 2D effects adjacent to the injection sites, the "pointwise'' distance $\nu$ and the "average'' distance $\zeta$ defined by Eqn.s~\ref{eqn:defn_functionals} are computed at the same conditions ($T$, $\tilde{p}_{O_{2}}$, $\tilde{k}_{f}^{0}$) in the frequency range of $10^{-3}\le\omega\le10^5$ rad/s along the symmetry axis $\Gamma_{2}$, Fig.~\ref{fig:var_AR}. In the first line we plot the case where the sample is very thick with respect to the horizontal dimension ($AR=125$), both the $\nu_{eon}(\tilde y, \omega) = \nu\left[ \mu_{eon, 1D}, \mu_{eon, 2D}, \tilde y\right]$ and the $\zeta_{ion}(\tilde y, \omega) = \zeta\left[ \mu_{ion, 1D}, \mu_{ion, 2D}, \tilde y\right]$ are extremely small and the adjacency between 1D and 2D impedance is near perfect. If we decrease $AR$ to 12.5, then the 1D and 2D solutions tend to be further apart with $\nu_e\approx 25\%$ and $\zeta_e$ up to $20\%$. The difference between the two further increases at $AR = 5$ where the difference between impedance spectra is significant.

\section{The Effect of Diffusivity Gradients}
\subsection{Extension of the Model}

Interface effects are one of the biggest sources of uncertainty in doped ionics because impurities in doped materials tend to segregate near interfaces and affect electro-catalytic processes, absorption and diffusivities near the affected interfaces. Many studies \cite{hauch:A619} \cite{Sadykov-05} \cite{Wilkes200312} have attempted to address these issues. However, to the authors' knowledge, no continuum model has addressed yet the relationship of these changes to polarization resistance nor to impedance spectra. In this part of the paper we intend to address the effects of non uniform diffusivities, which are localized near the interfaces, and which we imagine are due to impurity segregation at the exposed surface ($\Gamma_5$ in Fig.~\ref{fig:mat_domain}) and to the MIEC$|$metal interface ($\Gamma_4$)


We shall assume that diffusivities near the MIEC$|$Gas interface and MIEC$|$Metal interfaces have non-zero derivatives only along the $y$ direction. We further assume that diffusive effects are symmetric on both ends of the sample $y = \pm l_2$, hence do not affect our initial symmetry assumptions. Lastly we suppose that the functional form of the diffusivities are known in the MIEC and are given by:
\begin{equation}
D_m^\star = 1+\left(\frac{D_m^{SURF}}{D_m^{BULK}}-1\right) e^{-\frac{|l_c\tilde y\pm l_2|}{\lambda_m}}
\end{equation}
where $m$ can be either $eon$ or $ion$, and $\lambda_m$, the length scale of diffusive changes, is much smaller than $l_c$, the characteristic length-scale of the sample ($\lambda_m\ll l_c$). We stress again that the main assumptions are that the diffusivity gradients parallel to the interfaces are null and that the diffusivity gradients do not affect bulk properties of the material nor the defect chemistry. In other words, near-interface effects involve only diffusivities.

Under the same small perturbation assumptions we used above we can deduce that the equations that describe the impedance spectra behavior of ions and electrons are given by~\footnote{In order to ensure  linearity, we assume that $\left |D_k n^{(1)}\nabla \tilde \phi^{(1)}\right |\ll \left |D_k \nabla n^{(1)}\right | \approx \left |D_k \nabla \tilde \phi\right |$}:
\begin{subequations}
\begin{eqnarray}
&&n^{(1)} = \frac{\bar n}{\bar p }p^{(1)}\\
&&\displaystyle \frac{\tau_n}{\tau} \partial_{\tilde t}n^{(1)}+\nabla_{\tilde x}\cdot\left(-D_{eon}^\star\left(\nabla_{\tilde x} n^{(1)}-\nabla_{\tilde x} \tilde \phi^{(1)} \right) \right) = 0\\
&&\frac{\tau_p}{\tau} \partial_{\tilde t}p^{(1)}+\nabla_{\tilde x}\cdot\left(-D_{ion}^\star\left(\nabla_{\tilde x} p^{(1)}+2 \nabla_{\tilde x} \tilde \phi^{(1)} \right) \right) =0
\end{eqnarray}
\label{eqn:e-neutral_vary_diffusivity}
\end{subequations}
The sum of the Eqns.~\ref{eqn:e-neutral_vary_diffusivity} and their weighted difference lead to (see appendix \ref{appendix_B}):
\begin{subequations}
\begin{eqnarray}
\frac{\tau_n^\star}{\tau} \partial_{\tilde t}n^{(1)}+\nabla_{\tilde x}\cdot\left(-a_{11} \nabla_{\tilde x} n^{(1)}-a_{12} \nabla_{\tilde x}\tilde \phi^{(1)}  \right) = 0\\
\frac{\tau_\phi^\star}{\tau}\partial_{\tilde t}n^{(1)}+\nabla_{\tilde x}\cdot\left(-a_{21} \nabla_{\tilde x} n^{(1)}-a_{22}\nabla_{\tilde x} \tilde \phi^{(1)} \right) =0
\end{eqnarray}
\label{eqn:e-neutral_vary_diffusivity_shuffle}
\end{subequations}
where:
\begin{subequations}
\begin{eqnarray}
a_{11} &= \displaystyle \frac{D_{eon}^\star+\frac{\bar n}{4 \bar p}D_{ion}^\star}{1+\frac{\bar n}{4 \bar p}} ~;~
a_{12} &= \displaystyle \frac{D_{ion}^\star-D_{eon}^\star}{1+\frac{\bar n}{4 \bar p}} \\
a_{21} &= \displaystyle\frac{D_{ion}^\star-D_{eon}^\star}{1+\frac{4\bar p}{\bar n}} ~;~
a_{22} &= \displaystyle \frac{D_{eon}^\star+\frac{4\bar p}{\bar n}D_{ion}^\star}{1+\frac{4\bar p}{\bar n}}
\end{eqnarray}
\label{eqn:defn_coeffs_diffusivity}
\end{subequations}
\noindent The Eqn.s \ref{eqn:e-neutral_vary_diffusivity_shuffle} with appropriate boundary conditions, Eqn.s~\ref{eqn:BC_to_solve_FT}, are quasi-linear and hence can be Fourier transformed. In short they can be recast in weak form as in Eqns. \ref{eqn:to_solve_FT_weak}:



 
\begin{subequations}
\begin{eqnarray}
\begin{array}{lcl}
\displaystyle\omega \tau^\star_n \int_\Omega \hat n^{(1)}_{Im} m_{Re} \ud \tilde A &-& \displaystyle\int_\Omega a_{11} \nabla \hat n^{(1)}_{Re} \cdot \nabla m_{Re} \ud \tilde A -\int_\Omega a_{12} \nabla \hat \phi^{(1)}_{Re} \cdot \nabla m_{Re} \ud \tilde A+\ldots\\
\displaystyle&\ldots&\displaystyle+\int_{\Gamma_5} \tilde A_{n,2} \hat n^{(1)}_{Re} m_{Re} \ud \tilde A  - 4 \frac{\bar p}{\bar n}\int_{\Gamma_4} \partial_{\tilde y} \hat \phi ^{(1)}_{Re} m_{Re} \ud \tilde x =0
\end{array}\\
\begin{array}{lcl}
\displaystyle\omega \tau^\star_n \int_\Omega \hat n^{(1)}_{Re} m_{Im} \ud \tilde A &+& \displaystyle\int_\Omega a_{11} \nabla \hat n^{(1)}_{Im} \cdot \nabla m_{Im} \ud \tilde A+ \int_\Omega a_{12} \nabla \hat \phi^{(1)}_{Im} \cdot \nabla m_{Im} \ud \tilde A-\ldots \\
\displaystyle&\ldots&-\displaystyle\int_{\Gamma_5} \tilde A_{n,2} \hat n^{(1)}_{Im} m_{Im}  \ud \tilde x + 4 \frac{\bar p}{\bar n}\int_{\Gamma_4} \partial_{\tilde y} \hat \phi ^{(1)}_{Im}m_{Im} \ud \tilde x =0
\end{array}\\
\begin{array}{lcl}
\omega \tau^\star_\phi \int_\Omega \hat n^{(1)}_{Im} \psi_{Re} \ud \tilde A &-& \int_\Omega a_{21} \nabla \hat n^{(1)}_{Re} \cdot \nabla \psi_{Re} \ud \tilde A- \int_\Omega a_{22} \nabla \hat \phi^{(1)}_{Re} \cdot \nabla \psi_{Re} \ud \tilde A\\
&+&\int_{\Gamma_5} \tilde A_{\phi,2} \hat n^{(1)}_{Re} \psi_{Re} \ud \tilde x=0 \label{eqn:variation_phi_Re_diff}
\end{array}\\
\begin{array}{lcl}
\omega \tau^\star_\phi \int_\Omega \hat n^{(1)}_{Re} \psi_{Im} \ud \tilde A&+&  \int_\Omega a_{21} \nabla \hat n^{(1)}_{Im} \cdot \nabla \psi_{Im} \ud \tilde A + \int_\Omega a_{22} \nabla \hat \phi^{(1)}_{Im} \cdot \nabla \psi_{Im} \ud \tilde A\\
&-&\int_{\Gamma_5} \tilde A_{\phi,2} \hat n^{(1)}_{Im} \psi_{Im}\ud \tilde x=0 \label{eqn:variation_phi_Im_diff}
\end{array}
\end{eqnarray}
\label{eqn:to_solve_FT_weak_diff}
\end{subequations}

where:
\begin{eqnarray}
\tilde A_{n,2} = a_{11}\tilde A_n +a_{12}\tilde A_\phi \\
\tilde A_{\phi,2} = a_{21}\tilde A_n +a_{12}\tilde A_\phi \\
\end{eqnarray}

\noindent If we change the diffusivity of vacancies at the gas$|$ceria ($\Gamma_5$) and metal$|$ceria ($\Gamma_4$) interface by changing $\alpha_{ion}$, we need to adjust the $\tilde k_f^0$ as follows, in order to keep the same rate of injection $\dot \omega_{eon}^S$, Eqn.~\ref{eqn:chang_jaffe}:

\begin{equation}
\tilde k_f^{(0)} \left(\alpha_{ion}\right)=\frac{\left(\alpha_{ion}\right)_{\rm ref}}{\alpha_{ion}}\left(\tilde k_f^0\right)_{ref}
\end{equation}

\noindent Numerically we use the same approach described for the linear case but we need the error estimator to account for off-diagonal and space dependent parameters, Eqn.s \ref{eqn:defn_coeffs_diffusivity} (in the linear case $a_{11}= a_{22} =1$, $a_{12}= a_{21} =0$).

Finally we note that we assume that the model holds for length-scales just one order of magnitude greater that the lattice parameter \cite{zhan:A427}. This approximation can be justified heuristically using the work of Armstrong \cite{Armstrong} \cite{Horrocks_Armstrong}, which shows that deviations of the continuum drift-diffusion approach from atomistic models are usually small, even in cases where field effects are big.

\subsection{Results of the Model}

We first ran the model at steady state ($\omega =0$) with the objective  to analyze the $f_{surf} =\frac{R_{surf}}{R_{ion}^\perp}$ at $\omega =0$ for a wide array of parameters $\alpha_{eon}={D_{eon}^{SURF}}/ {D_{eon}^{BULK}}$ and $\alpha_{ion}={D_{ion}^{SURF}}/{D_{ion}^{BULK}}$, where $\alpha_{eon}= \alpha_{ion}$ and $\lambda_{eon} = \lambda_{ion}$ at varying $\tilde k_f^{(0)}$.
For reasonable fitted values (Tab.~\ref{tab:fitted_kf0}) and for a wide parameter set, we show that the polarization resistance is surface dominated making $f_{surf}\approx 1$ robustly. 

\begin{figure}
\scalebox{0.4}{\includegraphics{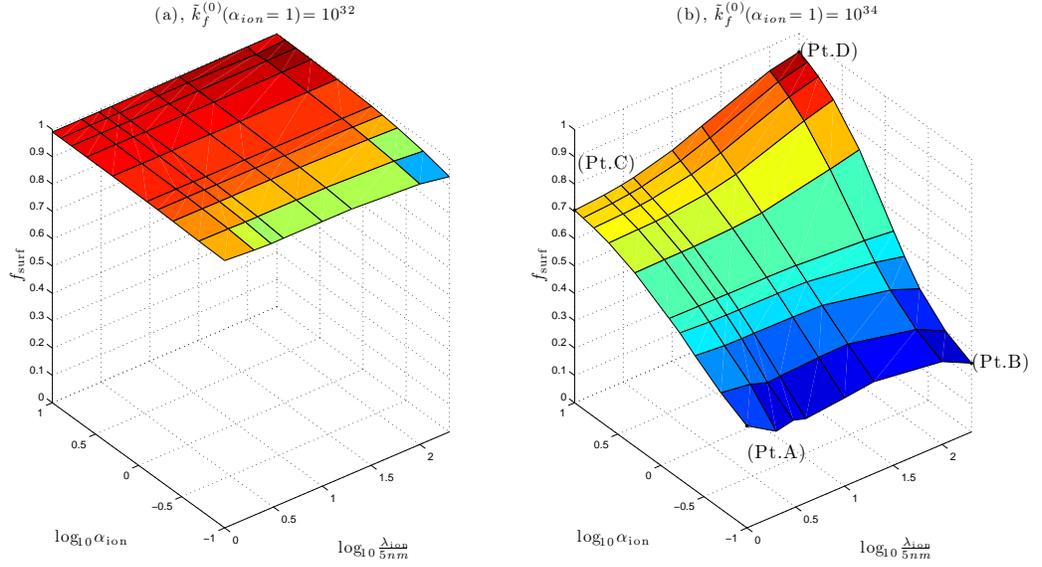}}
\caption{\label{fig:f_surf_D_same_dir} Depiction of $f_{surf}$ in the case $T=650^oC$ and $\tilde p_{O_2} = 10^{-25.32}$ as a function of the ratio between near interface and bulk diffusivity, $\alpha_{ion}= D_{ion}^{SURF}/D_{ion}^{BULK}$ and $\alpha_{eon}= D_{eon}^{SURF}/D_{eon}^{BULK}$ ($\alpha_{ion} = \alpha_{eon}$), and length scale of the diffusive gradient $\lambda_{ion} = \lambda_{eon}$, for $k_f^{(0)}=10^{32}$ (left panel) and $k_f^{(0)}=10^{34}$ (right panel).
}
\end{figure}

\begin{figure}
\scalebox{0.45}{\includegraphics{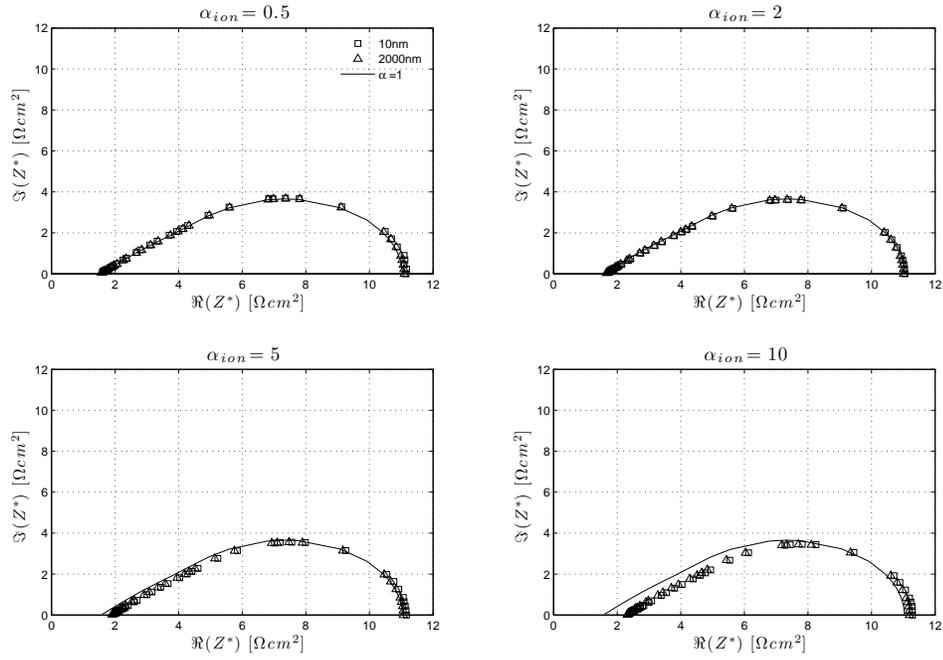}}
\caption{\label{fig:impedance_D_same_dir} Impedance of the sample under the conditions: $\tilde k_f^{(0)}=10^{32}$, $\tilde p_{O_2} =10^{-25.33}$ and $T = 650^oC$, where $\alpha_{eon} = \alpha_{ion}$ ($\alpha_m = D^{\rm SURF}_m/D^{\rm BULK}_m$) and $\lambda_{ion}= \lambda_{eon}$. The solid line represents the case where $\alpha_{ion} =1$, the triangles and the squares indicare respectively $\lambda_{ion} = 5nm$ and $\lambda_{ion} = 1\mu m$. Each panel corresponds to a different value of $\alpha_{ion}$. Only small deviations occur from the case $\alpha_{ion} =1$}
\end{figure}

If chemical reaction rates are "sufficiently" slow (e.g. $\tilde k_f^0\approx 10^{32}$) and if the sample is sufficiently thick, then the polarization resistance is dominated by surface effects in the linear case ($\alpha_{ion}=1$), corresponding to an absence of diffusive gradients at the exposed surface. If impurities are present at the exposed surface, diffusivities of charged species may change and hence one could argue that the polarization resistance is not surface-dominated. In order to address this point, we ran two limiting cases, one featuring "slow" chemistry ($\tilde k_f^0 (\alpha_{ion} =1)\approx 10^{32}$) and the other one at "fast" chemistry ($\tilde k_f^0(\alpha_{ion}=1)\approx 10^{34}$). We present the results of these calculations in Fig.~\ref{fig:f_surf_D_same_dir} where we plot $f_{surf}$ as a function of both $\alpha_{ion}=\alpha_{eon}$ and the diffusive gradients $\lambda_{ion} = \lambda_{eon}$. We notice from Fig.~\ref{fig:f_surf_D_same_dir}a that $f_{surf}$ is very close to  unity for two order of variation of surface-to-bulk diffusivity ratio $0.1\leq \alpha_{ion} \leq 10$ and for a wide span of diffusivity length-scales $5nm\leq \lambda_{ion}\leq 1 \mu m$. This indicates that if we perturb the the surface diffusivity up to one orders of magnitude with respect to its bulk value its impact on polarization resistance is minimal. The qualitative effect on the impedance is also small as shown for a variety of cases in Fig.~\ref{fig:impedance_D_same_dir}. 

If we choose a "fast" chemistry condition instead, e.g. $\tilde k_f^0 \approx 10^{34}$, the situation changes significantly from the base case ($\alpha_{ion} = 1$), Fig. ~\ref{fig:impedance_D_same_dir}b. In this Figure we focus on points A through D. (Pt. A), having $\alpha_{ion} = 0.1$ and $\lambda_{ion} = 5nm$, indicates that near surface diffusivities are an order of magnitude lower than their bulk value and this deviation is concentrated near the surface: in this case the polarization resistance is drift-diffusion dominated. If the diffusive length scale is increased to $\lambda_{ion} = 1\mu m$, while keeping $\alpha_{ion} = 0.1$, (Pt. B), the $f_{surf}$ will not decrease much further. Starting from (Pt.~A) we can move to (Pt.~C), where diffusivity gradients are sharp ($\lambda_{ion} = 5nm$) but the diffusivities at the surface are an order of magnitude greater than its bulk value. In this case, the $f_{surf}$ increases because of the increase in the bulk diffusivity. Going to (Pt.~C) to (Pt.~D) increases the length-scale of the diffusive effects leading in turn to bigger increase of $f_{surf}$.

We can summarize our findings as follows:
\begin{enumerate}
\item{if the rate of injection of electrons is sufficiently "small" (slow chemistry) and of the order of the fitted values reported in Tab.~\ref{tab:fitted_kf0}, then the diffusivity grandients localized at interfaces will affect little the polarization resistance and the impedance spectra;}
\item{if the chemistry is sufficiently fast, sharp changes in diffusivity can affect strongly not only the impedance behavior but also the polatization, in particular if the diffusivities increase sufficiently, strictly near the interfaces, the polarization effects will shift to be surface dominated, while a decrease is associated to drift-diffusion dominated polarization resistance.}
\end{enumerate}

\section{Concluding Remarks}

A general two-dimensional numerical framework has been developed for the coupled surface chemistry, electrochemistry and transport processes in mixed conductors based on the finite element method. As a specialized application of the framework, a time-dependent model was formulated based on first-principles for the AC impedance spectra (IS) of a samaria doped ceria (SDC) electrolyte with symmetric metal patterns on both sides, and the IS was simulated for typical fuel cell operation conditions in a uniform gas atmosphere ($H_2$, $H_2O$, $Ar$) at thermodynamic equilibrium using the small perturbation technique.

The validity of the model is demonstrated by fitting to experimental (1D) impedance spectra data of an SDC cell in literature, varying only the reaction rate at the SDC-gas interface. Excellent agreement ($\leq 2\%$ error) was obtained. We then numericallly investigated the influence of the variation of several parameters on the polarization resistance and the impedance spectra, especially within regimes not probable for the 1D studies. Our calculation shows that the 2D effect of cell thickness variation on the spectra becomes pronounced as the aspect ratio goes below a certain threshold (25 for this work); surface reaction dominates the polarization resistance when the injection rate at the SDC surface exposed to gas is sufficiently slow; sharp gradients in diffusion coefficient strongly influence both impedance behavior and polarization when surface chemistry is sufficiently fast. 

The discussions in this work provide useful insights into the correlation between materials properties of SDC and its applications in fuel cells, intensely studied by the solid oxide fuel cell researchers. In addition, the geometric capability (up to 3D) and high computation efficiency makes this numerical framework an ideal tool for the general study of mixed conductors. 

\section{Acknowledgments}
The authors gratefully acknowledge financial support for this 
work by the Office of Naval Research under grant N00014-05-1-0712. \\
The authors thank Prof. Fr\'ed\'eric Hecht for his valuable insight and support on 
Freefem++. 

\newpage 

\appendix
\section{Error Estimator and Refinement Strategy}
\label{appendix_A}

The local residual for $n_{Re}$, at a triangular element $K$ of the mesh, can be computed as follows \cite{Brenner-00}:

\begin{equation}
\begin{array}{lcl}
\eta_{k,~n_{Re}}&=&\displaystyle \int_K \left|\nabla\cdot\left(a_{11} \nabla \hat n_{Re, h}^{(1)}+a_{12} \nabla \hat \phi_{Re, h}^{(1)}\right) -\omega \tau_n^\star \hat n_{Im,h}^{(1)} \right| h^2+\left\llbracket a_{11}\frac{\partial \hat n_{Re, h}^{(1)} }{\partial\mathbf n} +a_{12}\frac{\partial \hat \phi_{Re, h}^{(1)} }{\partial\mathbf n}\right\rrbracket h_K^{1/2}\\
&+& \displaystyle\int_{\Gamma_5\cap K}\left| \tilde A_{n,2} \hat n_{Re,h}^{(1)} - \partial_{\tilde y} \hat n_{Re,h}^{(1)}\right|h^2+ \displaystyle\int_{\Gamma_4\cap K}\left|\partial_{\tilde y} \hat n_{Re}^{(1)} - 4 \frac{\bar p}{\bar n} \partial_{\tilde y} \hat n_{Re,h}^{(1)}\right| h^2+ \displaystyle\int_{(\Gamma_2\cup \Gamma_3)\cap K}\left|\partial_{\tilde x} \hat n_{Re}^{(1)} \right| h^2
\end{array}
\end{equation}

where $\left\llbracket a \right\rrbracket$ is the jump of the quantity $a$ across the faces of $K$, $h$ is a measure of the size $K$, while $h_K$ is the measure of the size of the sides of $K$. Similar residuals can be found for $n_{Im}^{(1)} $, $\phi_{Re}^{(1)} $, $\phi_{Im}^{(1)} $. Their sum $\sum_k r_k$ constitutes a reasonable local a posteriori error estimator. $\sum_k r_k$ is a weakly coercive upper bound for $a\Vert u \Vert_{L_2} - b\Vert \nabla u\Vert_{L_2}$ where $a$ and $b$ are constants and $u = \left(n_{Re}^{(1)}, n_{Im}^{(1)}, \phi_{Re}^{(1)}, \phi_{Im}^{(1)}\right)$

\section{Derivation of the Non-linear impedance Spectra Equations}
\label{appendix_B}

We start with the electro-neutral form of the drift-diffusion equations, where we assume that the diffusion coefficients normalized with respect to their bulk value $D_m^\star = D_m^{\rm SURF}/D_m^{\rm BULK}$:

\begin{subequations}
\begin{eqnarray}
\frac{\tau_n}{\tau} \partial_{\tilde t}n^{(1)}+\nabla_{\tilde x}\cdot\left(-D_{eon}^\star\left(\nabla_{\tilde x} n^{(1)}-\nabla_{\tilde x} \tilde \phi^{(1)} \right) \right) = 0 \label{eqn:appendix_B_DDn}\\
\frac{\bar n}{4 \bar p}\frac{\tau_p}{\tau} \partial_{\tilde t}n^{(1)}+\nabla_{\tilde x}\cdot\left(-D_{ion}^\star\left(\frac{\bar n}{4 \bar p}\nabla_{\tilde x} n^{(1)}+ \nabla_{\tilde x}\tilde\phi^{(1)} \right) \right) =0  \label{eqn:appendix_B_DDp}
\end{eqnarray}
\end{subequations}

We first sum the Eqn.s~\ref{eqn:appendix_B_DDn} and  \ref{eqn:appendix_B_DDp} and obtain:

\begin{equation}
\left(\frac{\tau_n}{\tau}+ \frac{\bar n}{4 \bar p}\frac{\tau_p}{\tau} \right)\partial_{\tilde t}n^{(1)}+\nabla_{\tilde x}\cdot\left(-\left(D_{eon}^\star+ \frac{\bar n}{4 \bar p}D_{ion}^\star\right)\nabla_{\tilde x} n^{(1)}-\left(D_{ion}^\star-D_{eon}^\star\right)\nabla_{\tilde x}\tilde \phi^{(1)} \right) = 0 
\label{eqn:appendix_B_eqn_n}
\end{equation}

Finally we multiply Eqn. \ref{eqn:appendix_B_DDp} by $\frac{4 \bar p}{\bar n}$ and sum to Eqn~\ref{eqn:appendix_B_DDn}:

\begin{equation}
\left(\frac{\tau_p}{\tau}- \frac{\tau_n}{\tau} \right)\partial_{\tilde t}n^{(1)}+\nabla_{\tilde x}\cdot \left( -\left( D_{ion}^\star - D_{eon}^\star\right)\nabla_{\tilde x} n^{(1)}-\left(D_{eon}^\star+\frac{4 \bar p}{\bar n}D_{ion}^\star\right)\nabla_{\tilde x} \tilde\phi^{(1)} \right) = 0 
\label{eqn:appendix_B_eqn_phi}
\end{equation}

From the \label{eqn:appendix_B_eqn_n} and \label{eqn:appendix_B_eqn_phi}, the Eqn.s~\ref{eqn:e-neutral_vary_diffusivity_shuffle} follow immediately and so do their coefficients given in Eqn.s~\ref{eqn:defn_coeffs_diffusivity} .

\newpage 
\bibliographystyle{rsc}
\bibliography{refs}

\newpage
\begin{table}
\begin{center}
\caption{\label{tab:geometric_values} Data for the domain geometry and background doping}
\begin{tabular}{c|c}
$W_1$ & $1.5 ~\mu m$ \\
$W_2$ & $2.5~\mu m$ \\ 
$l_2$ & $ 500~\mu m$ \\
$l_c$ & $10~\mu m$ \\
$B$ & $3.47\times10^{+27}\frac{\text{\#particles}}{m^3}$
\end{tabular}
\end{center}
\end{table}

\begin{table}
\begin{center}
\caption{\label{tab:physic_params} Temperature range and material constants for the simulations.}
\begin{tabular}{c|cccc}
$T$&$ 500^oC$ & $550^oC$ & $600^oC$ &$650^oC$ \\ \hline
$K_g$ & $5.059{\mathrm E}{+27}$& $4.814{\mathrm E}{+25}$ & $7.757{\mathrm E}{+23}$ & $1.944{\mathrm E}{+22}$ \\
$K_r$ & $5.008{\mathrm E}{-22}$& $2.263{\mathrm E}{-20}$ & $6.610{\mathrm E}{-19}$ & $1.340{\mathrm E}{-17}$\\
$u_{eon}$ $\left[ \frac{m^2}{ V^2 s}\right]$  &$4.762{\mathrm E}-8$ &$6.257{\mathrm E}{-8}$ &$6.873{\mathrm E}{-8}$ & $8.123{\mathrm E}{-8}$\\
$u_{ion}$ $\left[ \frac{m^2}{ V^2 s}\right]$& $1.166{\mathrm E}{-9}$ & $2.070{\mathrm E}{-9}$ & $3.359{\mathrm E}{-9}$ & $4.936{\mathrm E}{-9}$
\end{tabular}
\end{center}
\end{table}

\begin{table}
\caption{\label{tab:definitions_1D} Definitions of the terms in the 1D model}
\begin{tabular}{c|cc}
$R_{ion}^\perp$ & Measured\\
$R_{eon}$ & ${2 l_2}/{\sigma_{eon}}$\\
$R_{ion}$ & ${2 l_2}/{\sigma_{ion}}$\\
$R_{0}$ & ${1}/\left({1/R_{eon}+1/\left( R_{ion}+2R_{ion}^\perp\right)}\right)$\\
$R_\infty$ & ${1}/\left({1/R_{eon}+1/R_{ion}}\right)$\\
$C_{chem}$ & $\frac{e^2}{k_b T}{2 l_2}/\left({1/(z_{eon}^2 c_{eon}^{(0)})+1/(z_{ion}^2 c_{ion}^{(0)})}\right)$\\
$\tilde D$ &  $4 l_2^2/\left(\left(R_{ion}+R_{eon}\right)C_{chem}\right)$ \\
$s$ & $\sqrt{{i4 \omega l_2^{2}}/({4\tilde{D}})}$
\end{tabular}
\end{table}

\begin{table}
\caption{\label{tab:fitted_kf0} Fitted values of $\tilde k_f^0 = A \tilde p_{O_2}^\alpha$, $95\%$ confidence interval}
\begin{tabular}{c|cccccc}
 $T~[^oC]$&$\log_{10}\bar A$&$\log_{10}\varepsilon_A$&$\bar \alpha$&
 $\varepsilon_\alpha$ &$R^2$&$\sigma$\\
\hline
$500$& 32.48 & 0.150 & 0.05349 & 0.1655 & -0.0439 & 0.1577 \\
$550$& 32.10 & 0.045 & 0.04160 & 0.0482 &  0.7622 & 0.04589 \\
$600$& 32.02 & 0.055 & 0.06674 & 0.0637 &  0.5378 & 0.06067 \\
$650$& 31.95 & 0.055 & 0.05596 & 0.0623 &  0.4981 &  0.05938
\end{tabular}
\end{table}

\newpage 

\begin{figure}
\scalebox{0.50}{\includegraphics{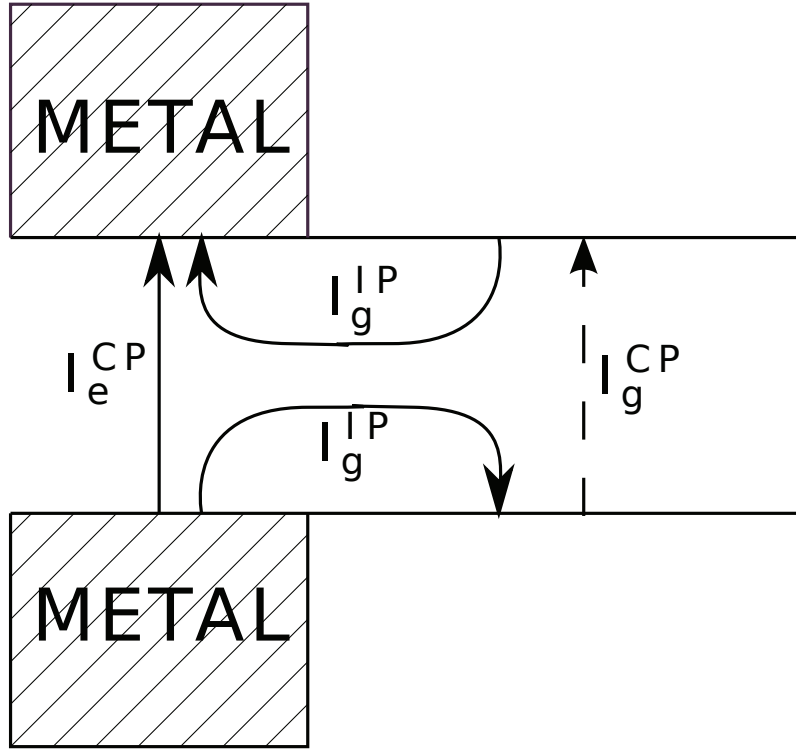}}
\caption{\label{fig:schematics_fluxes}Depiction of the currents in the MIEC. The superscript $CP$ indicates cross-plane current and the superscript  $IP$ means in-plane currents. The subscript $g$ indicates that the the flux is due to electrochemical reactions at the gas$|$ceria interface, while the subscript $e$ is for electrode to electrode current. We notice we will have four currents: one, the cross-plane electron flux $I_e^{CP}$ from the bottom to the top electrode, two the cross-plane ionic flux from top to bottom gas$|$ceria interface $I_g^{CP}$ and the in-plane electronic fluxes $I_g^{IP}$ from the gas$|$ceria interfaces to the electrodes.}
\end{figure}

\begin{figure}
\scalebox{0.5}{\includegraphics{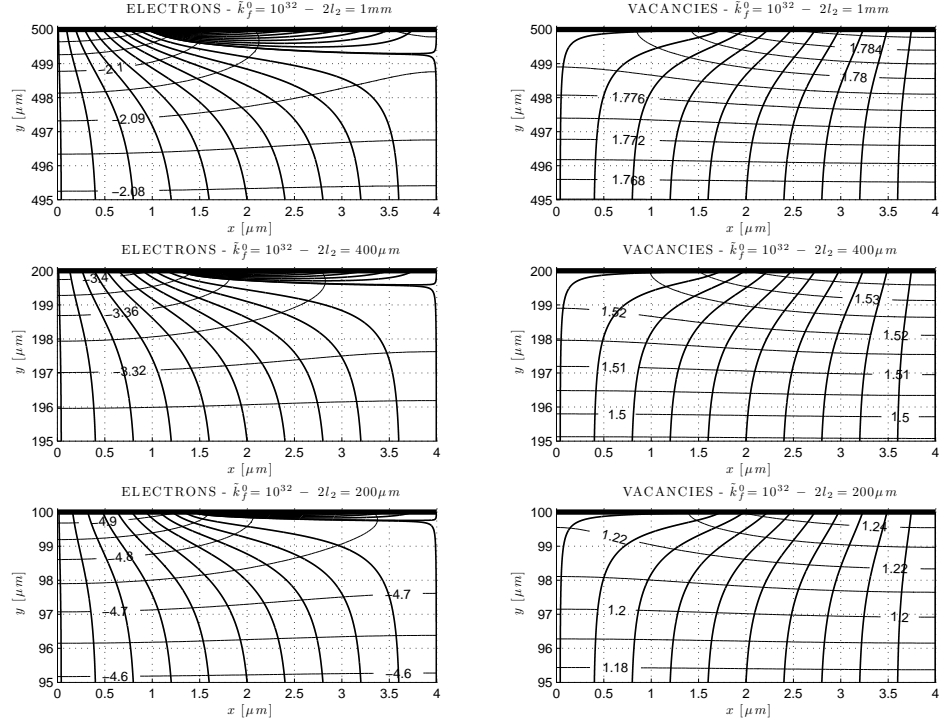}}
\caption{\label{fig:var_AR_omega_0} Potentials and current lines under small bias excitation, i.e. impedance at $\omega=0$, at $T=650^oC$ and $\tilde p_{O_2}=10^{-25.33}$. The $\hat \mu_{eon}$ (left column) and $\hat \mu_v$ (right column) along with their current lines are plotted. Each row corresponds to a different thickness. As $l_2$ decreases (from top to bottom row) the area affected by surface reactions thins out; this phenomenon relates to an increase of the polarization resistance.}
\end{figure}

\end{document}